\documentclass[10pt,amsmath,amssymb,twocolumn,superscriptaddress,floatfix,nofootinbib,aps]{revtex4-1}

\usepackage{graphicx}
\usepackage{dcolumn}
\usepackage{bm}
\usepackage{setspace}
\usepackage{amsmath}
\usepackage{enumerate}
\usepackage{tabularx}
\usepackage{wrapfig}
\usepackage{ifthen}
\usepackage[utf8]{inputenc}
\usepackage{color}

\usepackage[breaklinks=true,colorlinks=true]{hyperref}
\usepackage{breakurl}

\def\simge{\mathrel{
     \rlap{\raise 0.511ex \hbox{$>$}}{\lower 0.511ex \hbox{$\sim$}}}}
\def\simle{\mathrel{
     \rlap{\raise 0.511ex \hbox{$<$}}{\lower 0.511ex \hbox{$\sim$}}}}
\def\be{\begin{equation}}
\def\ee{\end{equation}}

\newcommand{\lr}[1]{ \left( #1 \right) }

\newcommand{\Tr}{ {\rm Tr} \, }

\newcommand{\calz}{{\cal Z}}

\newcommand{\beq}{\begin{eqnarray}}
\newcommand{\eeq}{\end{eqnarray}}

\DeclareMathOperator{\const}{const}

\graphicspath{{./figures/}}

\begin{document}
\sloppy

\title{Hybrid-Monte-Carlo study of competing order in the extended fermionic Hubbard model on the hexagonal lattice}

\author{Pavel~Buividovich}
\email{Pavel.Buividovich@physik.uni-regensburg.de}
\affiliation{Institut f\"ur Theoretische Physik, Universit\"at Regensburg, 93053 Regensburg, Germany}

\author{Dominik~Smith}
\email{Dominik.D.Smith@theo.physik.uni-giessen.de}
\affiliation{Institut f\"ur Theoretische Physik, Justus-Liebig-Universit\"at, 35392 Giessen, Germany}

\author{Maksim~Ulybyshev}
\email{Maksim.Ulybyshev@physik.uni-wuerzburg.de}
\affiliation{Institut f\"ur Theoretische Physik, Julius-Maximilians-Universit\"at,
   97074 W\"urzburg, Germany}

\author{Lorenz von Smekal}
\email{Lorenz.Smekal@physik.uni-giessen.de}
\affiliation{Institut f\"ur Theoretische Physik, Justus-Liebig-Universit\"at, 35392 Giessen, Germany}

\date{\today}
\begin{abstract}
\noindent 
Using first-principle Hybrid-Monte-Carlo (HMC) simulations, we carry out an unbiased study of the competition between spin-density wave (SDW) and charge-density wave (CDW) order in the extended Hubbard model on the two dimensional hexagonal lattice at half filling. We determine the phase diagram in the space of on-site and nearest-neighbor couplings $U$ and $V$ in the region $V<U/3$, which can be simulated without a fermion sign problem, and find that a transition from semimetal to a SDW phase occurs at sufficiently large $U$ for basically all $V$. Tracing the corresponding phase boundary from $V=0$ to the $V=U/3$ line, we find evidence for critical scaling in the Gross-Neveu universality class for the entire boundary. With rather high confidence we rule out the existence of the CDW ordered phase anywhere in the range of parameters considered. We also discuss several improvements of the HMC algorithm which are crucial to reach these conclusions, in particular the improved fermion action with exact sublattice symmetry and the complexification of the Hubbard-Stratonovich field to ensure the ergodicity of the algorithm. 
\end{abstract}
\maketitle

\section{Introduction}
\label{sec:Intro}

By now experimental \cite{Elias:12:1} and numerical \cite{Buividovich:13:5,Smekal:1403.3620} studies have firmly established that free suspended graphene is a semimetal. Applications in semiconductor electronics, however, require that a sizable energy gap should be opened in the band structure of graphene while preserving the extremely high carrier mobility \cite{Geim:07:1,Lanzara:PhysicsViewpoint}. 

This problem has motivated an active research on artificially modified graphene and graphene-like materials which might support gapped phases. New experimental techniques to control the microscopic interaction parameters are being rapidly developed. Ideas being discussed range from mechanically strained graphene \cite{Assaad:15:1,PhysRevB.96.155114} via 2D materials with hexagonal lattices such as phosphorene \cite{Liu:14:1}, silicene and germanene \cite{Cahangirov:09:1} to ``artificial graphene'' in optical lattices \cite{Tarruell:1111.5020}. Even more exotic materials, such as 3D Dirac semimetals \cite{PhysRevB.97.155122} or 2D semi-Dirac semimetals, which exhibit a dispersion relation which is linear along one momentum component but quadratic along the other one, are being considered \cite{PhysRevB.95.075129}. In many cases such systems can be described in terms of the extended Hubbard model on the hexagonal graphene lattice with nearest-neighbor hoppings and on-site and nearest-neighbor inter-electron interactions.

The hexagonal Hubbard model with varying on-site repulsion $U$ and nearest $V_1$ and next-to-nearest-neighbor $V_2$ interactions has been predicted to host a large variety of gapped phases with spontaneously induced order. These include anti-ferromagnetic (AF) spin-density wave (SDW) and charge-density wave (CDW) phases \cite{Sorella:1992,Semenoff:84:1,Herbut:cond-mat/0606195,Semenoff:1204.4531,Gracey:1801.01320}, topological insulators \cite{Raghu:07:1}, and spontaneous Kekul\'e distortions \cite{Mudry:cond-mat/0609740,PhysRevB.96.115132}. Even coupled spin-charge-density-wave phases as discussed for ultracold atoms in optical lattices \cite{Makogon:1007.0782} might occur, in principle. 
A detailed quantitative understanding of the phase diagram in the space of $U$, $V_1$ and $V_2$ couplings is desirable to guide experimental searches for non-trivial electronic ordered phases.\footnote{Realistic materials often exhibit non-zero interaction parameters at even larger distances (such as e.g. graphene, in which the bare interaction potential includes an unscreened Coulomb tail \cite{Wehling2011}). Renormalization group studies however show that these can be marginally relevant couplings, which may or may not be absorbed into the short-range interactions close to a phase transition (this was discussed for graphene in Refs. \cite{Herbut:cond-mat/0606195,Herbut2009,Herbut2009:2}).}

A reasonably good description of the expected phase structure is obtained from various semi-analytic methods, such as self-consistent random phase approximation \cite{Honerkamp:2017kmq} or a variational Hamiltonian approach \cite{Semenoff:1204.4531} and from ab-initio simulations using determinantal quantum Monte Carlo (DQMC) \cite{PhysRevB.89.205128,Assaad:1304.6340}.  Large-$N$ renormalization group fixed-point analysis reveals a complex structure of fixed points, depending on the number of fermion flavors. In the $V_2 =0$ plane of on-site $U$ and nearest-neighbor repulsion $V \equiv V_1$, it
is able to describe the universal behavior near a tentative multicritical point at which semimetal, CDW and SDW phases meet \cite{Herbut:cond-mat/0606195}. From an $\epsilon$-expansion around three spatial dimensions it was concluded in Ref.~\cite{PhysRevB.92.035429} that this point should be multicritical also in the case of graphene, with $N=2$, and that the behavior around this point should be dominated by the same chiral Heisenberg Gross-Neveu universality class that is also expected to describe the semimetal to SDW transition for smaller values of $V$. The latest large-$N$ results for the corresponding critical exponents are reported in Ref.~\cite{Gracey:1801.01320}. While there is convincing agreement between $\epsilon $-expansion and large-$N$ results for the universal properties of effective low-energy theories within this class close to the upper critical dimension, or for sufficiently large $N$, the situation for the two spatial dimensions and $N=2$ as relevant here appears to remain less clear. The functional renormalization group study of Ref.~\cite{PhysRevB.93.125119} for example predicts for $N=2$ a triple point where three first-order transition lines between semimetal, SDW and CDW phases meet in $U-V$ plane of the extended Hubbard model. 

In contrast to on-site repulsion $U$, a nearest-neighbor interaction $V$ acts equally between both spin components and therefore energetically favors CDW order. Moreover, because of the coordination number three, the interaction energies of on-site repulsion $U$ in the SDW ground state and nearest-neighbor repulsion $V$ in the CDW ground state are the same when $V = U/3$, and one thus expects a first-order phase transition with coexistence at sufficiently low temperatures along this $V= U/3$ line in the strong-coupling limit. In fact, it is possible to prove analytically that the Dyson-Schwinger equations in the static approximation, self-consistently including frequency independent screening beyond Hartree-Fock, are equivalent for CDW and SDW order along this line, i.e.\ that their solutions are in an exact one-to-one correspondence. Moreover, the free energies from the corresponding 2PI-effective action are the same in both gapped phases and the transition between the two must be discontinuous \cite{Kleeberg2018}. The Hartree-Fock phase diagram with the same qualitative behavior was presented in \cite{Buividovich:16:4}.

In this work we study the phase diagram of the extended Hubbard model on the hexagonal graphene lattice in the space of on-site repulsion $U$ and nearest-neighbor interaction $V$ using first-principle Monte-Carlo simulations. We use the Hybrid-Monte-Carlo (HMC) algorithm \cite{DeGrandDeTarLQCD,MontvayMuenster,Buividovich:16:1}, which is mainly used for lattice QCD simulations, but also gains increased popularity in recent years as a tool for condensed matter physics \cite{Hands:08:1,Hands:10:1,Hands:11:1,DelDebbio:96:1,Lahde:09:1,Lahde:09:2,Lahde:09:3,Lahde:11:1,Rebbi:12:1,Buividovich:13:5,Smekal:1403.3620,Assaad:1708.03661,Ulybyshev:2015opa,Ulybyshev:2017szp,PhysRevB.94.085421,PhysRevB.95.165442,PhysRevB.94.245112,Luu:2015gpl,Berkowitz:2017bsn,PhysRevB.96.165411,PhysRevB.96.205115}. As compared with our previous HMC simulations of graphene, the simulation algorithm used in this work includes several essential improvements:
\begin{itemize}
 \item Fermionic lattice action with exact sublattice (chiral) symmetry \cite{Buividovich:16:4}, which allows to make the discrete time step about an order of magnitude larger than for the straightforward first-order discretization.
 \item Complexified fields in the bosonic action which allow the molecular dynamics to penetrate the potential barriers due to zeros of the fermion determinant \cite{Assaad:1708.03661,Ulybyshev:1712.02188}.
 \item Efficient non-iterative Schur complement solver which significantly speeds up the simulations \cite{Buividovich:18:1}.
\end{itemize}

Using infinite-volume extrapolations of order parameters
and finite-size scaling, we are able to locate the boundary between the semimetal and the antiferromagnetic SDW phases, which shifts with $V$ towards larger critical values of $U$ as compared to the $V=0$ result $U_c \simeq 3.8 \kappa$ for pure on-site interactions obtained using DQMC with ground-state projection \cite{Assaad:1304.6340}. This shift has been observed previously in another DQMC study of the $U-V$ phase diagram \cite{PhysRevB.89.205128} but we find the effect to be much stronger, possible due to the dynamical cluster approximation which was employed in the previous study. At current precision our results for the squared spin per sublattice are consistent with critical scaling in the chiral Heisenberg Gross-Neveu universality class. Furthermore, with rather high confidence we exclude the existence of CDW and ferromagnetic phases in the parameter region with $V < U/3$, in which our HMC simulations have no sign problem. We point out that the use of the complexified Hubbard field is essential to reach this conclusion, as otherwise the presence of impenetrable potential barriers in configuration space produces a false signal also for CDW order, whenever the system exhibits SDW order. 

\section{Numerical setup}
\label{sec:Setup}

The algorithm used in this work is based on the formalism originally developed in Refs.~\cite{Rebbi:11:1,Rebbi:12:1} and has been described extensively, e.g. in Refs. \cite{Smekal:1403.3620,Ulybyshev:1712.02188,Korner:2017qhf,Buividovich:13:5}. We review the essential background in this section and highlight recent novel developments such as the improved fermion action, the use of the Schur solver and the complexification of the auxiliary Hubbard-Stratonovich fields. 

The starting point is the Hubbard-Coulomb Hamiltonian:
\beq 
\label{eq:tightbinding}
\hat{\mathcal{H}} = - \kappa \sum_{\left \langle x, y \right \rangle, \sigma} ( \hat{c}^{\dagger}_{x,\sigma} \hat{c}_{y,\sigma} + \text{h.c.}) + \frac{1}{2}\sum_{x,y}  \hat{\rho}_x V_{xy}\hat{\rho}_y .
\eeq
Here $\kappa$ is the hopping parameter, $\langle x, y\rangle$ denotes nearest-neighbor sites, $\sigma = \uparrow, \downarrow$ labels spin components and $\hat{\rho}_x =\hat{c}^{\dagger}_{x,\uparrow} \hat{c}_{x,\uparrow}+\hat{c}^{\dagger}_{x,\downarrow} \hat{c}_{x,\downarrow}-1$ is the electric charge operator.
The creation- and annihilation operators satisfy the anticommutation relations $\{ \hat{c}_{x,\sigma}, \hat{c}^{\dagger}_{y,\sigma'} \}= \delta_{x,y} \delta_{\sigma, \sigma'}$. In this work, the interaction is fully specified by on-site ($U \equiv V_{00}$) and nearest-neighbor ($V \equiv V_{01}$) couplings, which are treated as free parameters. HMC is applicable for positive-definite matrices $V_{xy}$, which leads to the restriction $V<U/3$ for a $2D$ hexagonal lattice. 

The basis of HMC is the functional integral representation of the grand-canonical partition function $\calz=\Tr e^{-\beta \hat{\mathcal{H}}}$, in which operators are replaced by fields. Thermodynamic averages of observables $\langle \hat{O}\rangle= \frac{1}{\calz}\Tr(\hat{O}e^{-\beta \hat{\mathcal{H}}})$ are then obtained from measurements on a representative set of field configurations, generated in proportion to their weight in the equilibrium ensemble. The Hamiltonian (\ref{eq:tightbinding}) is free of a fermion sign problem (where the measure of the functional integral is complex or of indefinite sign, which prevents importance sampling) on a bipartite lattice at half-filling after introducing hole operators with a sublattice-dependent phase for the spin-down electrons, i.e. after applying the transformation
\begin{eqnarray}
\hat{c}_{x, \uparrow}, \hat{c}^{\dagger}_{x, \uparrow} &\to & \hat{a}_x, \hat{a}^{\dagger}_x,
\nonumber \\ 
\label{eq:trafo1}
\hat{c}_{x, \downarrow}, \hat{c}^{\dagger}_{x, \downarrow} &\to & \pm \hat{b}^{\dagger}_x, \pm \hat{b}_x,
\end{eqnarray}
where the signs in the second line alternate between the two sublattices. This also leads to $\hat{\rho}_x = \hat{a}^{\dagger}_x \hat{a}_x - \hat{b}^{\dagger}_x \hat{b}_x$.

To derive the functional integral, we start with a symmetric Suzuki-Trotter decomposition which yields 
\begin{eqnarray}
\calz &\approx &\Tr \left(
\prod_{i=1}^{N_\tau}
e^{-\delta_{\tau}(\hat{\mathcal{H}}_{0} +\mathcal{H}_{\text{int}})} \right) \notag\\ &=&
\Tr \left( e^{-\delta_{\tau} \hat{\mathcal{H}}_{0}} e^{-\delta_{\tau} \hat{\mathcal{H}}_{\text{int}}} e^{-\delta_{\tau} \hat{\mathcal{H}}_{0}} \dots \right) + O(\delta^2_{\tau}),
\label{eq:Trotter}
\end{eqnarray}
where the exponential is factorized into $N_\tau$ terms and the kinetic $\hat{\mathcal{H}}_{0}$ and interaction $\hat{\mathcal{H}}_{\text{int}}$ contributions are separated. This introduces a finite step size $\delta_{\tau} = \beta/N_{\tau}$ in Euclidean time and a discretization error $O(\delta^2_{\tau})$. The separation of $\hat{\mathcal{H}}_{0}$ and $\hat{\mathcal{H}}_{\text{int}}$ in the second line arises from symmetrized second-order approximants for each factor inside the trace in the first line and effectively doubles the number of time slices. The advantage of this expansion will become clear below. 

The four-fermion terms appearing in $\hat{\mathcal{H}}_{\text{int}}$ should now be converted into bilinears. This step is essential, since we can then explicitly integrate out the fermionic operators. This is achieved by Hubbard-Stratonovich (HS) transformation
\begin{eqnarray}
\label{continuous_HS_imag}
  e^{-\frac{\delta_\tau}{2}\sum_{x,y} V_{x,y} \hat \rho_x \hat \rho_y} \cong \int D \phi \,
  e^{- \frac{1}{2\delta_\tau} \underset{x,y}{\sum} \phi_x V^{-1}_{xy} \phi_y} e^{i \underset{x}{\sum} \phi_x \hat \rho_x}, 
\end{eqnarray}
at the expense of introducing a bosonic auxiliary field $\phi$ (``Hubbard field''). Eq.~(\ref{continuous_HS_imag}) is applied once to each timeslice, leading to $\phi\equiv\phi_{x,t}$. Note that this form of the HS transformation, using a non-compact continuous Hubbard field and a purely imaginary exponent in the rightmost term, is only one of many possibilities. 
At the end of this Section we will discuss another variant, which is used to prevent violations of ergodicity. 

To compute the trace in the fermionic Fock space (with anti-periodic boundary conditions) one uses the identity   
\begin{align}
\label{fermionic_identity}
 &\Tr\left( e^{-\hat{A}_1 } e^{-\hat{A}_2 } \ldots e^{-\hat{A}_n } \right)
 = \nonumber \\ &=
 \det\left(
  \begin{array}{cccc}
     1          & -e^{-A_1} & 0        & \ldots  \\
     0          & 1        & -e^{-A_2} & \ldots  \\
        \vdots  &          & \ddots          &         \\
      e^{-A_n} & 0        & \ldots   & 1       \\
  \end{array}
 \right)
 = \nonumber \\ &=
 \det\left(1 + e^{-A_1} e^{-A_2} \ldots e^{-A_n} \right)
  ,
\end{align}
for even $n$, where $\hat{A}_k = \lr{A_k}_{ij} \hat{c}^{\dag}_i \hat{c}_j$ are the fermionic bilinear operators and $A_k$ (without hat) contain matrix elements in the single-particle Hilbert space. The expressions (\ref{fermionic_identity}) are derived in Refs. \cite{Hirsch:85:1,Blankenbecler:81:1,MontvayMuenster} and are also the core of the determinantal Quantum-Monte-Carlo simulations following Blankenbecler, Scalapino and Sugar (BSS). Applying (\ref{fermionic_identity}) to the expression (\ref{eq:Trotter}), we obtain
\begin{eqnarray}
\label{eq:func_int}
  \calz &=& \int D \phi \,|\det M( \phi)|^2
  e^{-S_\phi},\,\\ S_\phi &=&\frac{1}{2\delta_\tau} \sum_{x,y,t} \phi_{x,t} V^{-1}_{xy} \phi_{y,t}\,,
\end{eqnarray}
which fulfills the basic requirements of HMC, in the sense that the integrand in (\ref{eq:func_int}) can be interpreted as a classical probability density for the Hubbard field. The fermion matrix is given by
\begin{align}
\label{eq:fermionmatrix}
 &M(\phi) 
 = \nonumber \\ &=
 \left(
  \begin{array}{cccccc}
     1          & -e^{-\delta_\tau h} & 0        & 0 & 0 &\ldots  \\
     0          & 1        & -e^{i \phi_{1}} & 0 & 0 & \ldots  \\
     0          & 0        & 1                & -e^{-\delta_\tau h} & 0 &    \ldots\\
     0          & 0 & 0     & 1                & -e^{ i \phi_{2}} &    \ldots\\
         \vdots & & & & \ddots &                               \\
     e^{i \phi_{N_\tau}} & 0  & 0 &       & \ldots   & 1       \\
  \end{array}
 \right) ,
\end{align}
where $h$ denotes the single-particle tight-binding hopping matrix and we use the short-hand notation $e^{i \phi_t} \equiv \textrm{diag}\left( e^{i \phi_{x,t}} \right)$ for the exponentiated Hubbard-Stratonovich fields which are packed into a diagonal matrix and interpreted as operators on the single-particle Hilbert space. $|\det M( \phi)|^2$ appears in (\ref{eq:func_int}) since after the transformation (\ref{eq:trafo1}) the fermionic matrices for spin-up and spin-down electrons are $M$ and $M^\dagger$, respectively. The doubling of time-slices is manifest in (\ref{eq:fermionmatrix}) and the Hubbard fields appear only in the even ones. Note that in the fermion matrix (\ref{eq:fermionmatrix}) the time derivative $\partial_{\tau}$ is discretized as a forward finite difference of the form $\psi_{t+1} - \psi_{t}$ which does not suffer from the fermion doubling problem at the expense of not being anti-Hermitian. Since only the combination $M M^{\dag}$ enters in the path integral weight, this does not cause any problems in our simulations. However, the sublattice (pseudospin) and the spin (flavor) degrees of freedom are both needed for this positivity so that, together with the two Dirac cones, the total number of 8 massless fermionic excitations per Brillouin zone in the present setup is actually exactly the same as that on a cubic lattice with the usual doublers.

 Moreover, since the spatial lattice spacing is fixed for graphene, we can smoothly take the time continuum limit $\delta_{\tau} \rightarrow 0$ without encountering any ultraviolet divergences. In essence, due to finite spatial lattice spacing graphene can be treated as a quantum-mechanical system where UV divergences do not appear.

The fermion matrix $M(\phi)$ in (\ref{eq:fermionmatrix}) differs from the one used in several previous HMC studies of fermionic Hubbard models on the hexagonal lattice \cite{Buividovich:12:1,Smekal:1403.3620,Korner:2017qhf}, and is closer to the form used in BSS QMC simulations. The difference arises entirely from the way the non-interaction tight-binding hopping term is discretized in the derivation of the lattice action in the partition function. Roughly speaking, inserting complete sets of fermionic coherent states $|\xi\rangle $ (with $c_i|\xi \rangle = \xi_i | \xi\rangle$) in between all factors in Eq.~(\ref{eq:Trotter}), the previously used linear action is obtained from matrix elements
\begin{align}
\langle\bar\xi| e^{-\delta h_{ij} c_i^\dagger c_j } |\xi\rangle &=
e^{\bar\xi_i \xi_i} \, \left( 1 -  \delta  h_{ij} \bar\xi_i\xi_j  + \mathcal O(\delta^2) \right)   \nonumber \\
&= e^{\bar\xi_i \xi_i - \delta  h_{ij} \bar\xi_i\xi_j } + \mathcal O(\delta^2) \, ,
\end{align}
where the error is due to neglected normal-ordering terms that arise at the order $\delta^2$ when expanding the exponential. These can be summed by instead using the formula,
\begin{align}
\langle\bar\xi| e^{-\delta h_{ij} c_i^\dagger c_j } |\xi\rangle = e^{\bar\xi_i (e^{- \delta  h}) _{ij} \xi_j } \, .
\end{align}
The same summation of normal-ordering terms was already used in the previous studies to derive the compact Hubbard-field interaction $\propto e^{i \phi_t}$ in the fermion matrix. Here we also use it for the free tight-binding hamiltonian to derive the fully exponential action with the fermion matrix in Eq.~(\ref{eq:fermionmatrix}). The linearized action of the previous studies thus corresponds to 
expanding the blocks $e^{-\delta_\tau h}$ in the fermion matrix $M(\phi)$ to linear order in $\delta_\tau$ again, which amounts to replacing them by $1 - \delta_{\tau} h$. The main disadvantage of this linearized formulation is that the leading discretization errors generate a strong explicit breaking of the spin rotational symmetry, which is only suppressed at very large $N_\tau$ as observed in \cite{Buividovich:16:4}. In practice, using the fermion matrix (\ref{eq:fermionmatrix}) with exact sublattice symmetry allows us to use a Euclidean time step $\delta_{\tau}$ which is $\sim 10$ times larger than that for the first-order discretization at the same level of discretization errors.

The origin of this asymmetry lies in the mixing of spin and sublattice symmetries after applying the transformation (\ref{eq:trafo1}). One can see this by defining a generator $\Sigma_{xy}$ of the sublattice symmetry in the single-particle Hilbert space, whose matrix elements are 
non-vanishing only for $x = y$, and are $+1$ on one sublattice and $-1$ on the other. In absence of mass terms, the single-particle hopping matrix $h$ then satisfies the identity $\Sigma h \Sigma = - h$. Analogous to the action of the $\gamma_5$ matrix on the Dirac Hamiltonian, this amounts to exchanging positive-energy and negative-energy states. These are equivalent, however, by virtue of the particle-hole symmetry of the bipartite lattice. 
The above identity implies $\Sigma e^{-\beta h} \Sigma = e^{\beta h}$ and the partition function thus remains invariant under this symmetry. If one discretizes Euclidean time into $N_\tau$ intervals of size ${\delta_\tau}$, and at the same time expands the single-particle transfer matrix $e^{-\delta_\tau h} \approx 1 - {\delta_\tau} h $ this no longer holds, since $\Sigma \lr{1 - {\delta_\tau} h} \Sigma = \lr{1 + {\delta_\tau} h} \neq \lr{1 - {\delta_\tau} h}^{-1}$. In other words, the particle transfer matrix is no longer the inverse of the hole transfer matrix. A particle propagating backwards in time is no-longer equivalent to a hole, and thus the combined particle-hole and sublattice symmetries are violated by corrections of order $\delta_\tau$.
Since particles and holes were identified with spin components in Eqs.~(\ref{eq:trafo1}), this violation translates into one of the spin symmetry.

In contrast, the fermion matrix (\ref{eq:fermionmatrix}) has an exact sublattice-particle-hole symmetry even at finite $\delta_\tau$ and in the presence of the fluctuating Hubbard fields \cite{Buividovich:16:4}. The price we pay is that, while $1 - \delta_{\tau} h $ is a sparse matrix, $e^{-\delta_{\tau} h}$ is not. This makes iterative inversion methods such as the standard conjugate-gradient solver rather inefficient for the inversion of a fermion matrix of the form in (\ref{eq:fermionmatrix}). The situation here is analogous to lattice QCD simulations with exactly chiral fermions, where exact chiral symmetry can only be preserved with a non-local action \cite{Neuberger:98:1}. HMC simulations based on 
Eq.~(\ref{eq:fermionmatrix}) have become feasible only recently with the development of a novel non-iterative solver based on Schur decomposition \cite{Buividovich:18:1}. This solver also tremendously speeds up the calculation of observables, especially those requiring the inversion of $M(\phi)$ on multiple right-hand side vectors at fixed $\phi$. All the results in this work were obtained using this novel solver, which we briefly describe in Appendix \ref{apdx:Schursolver} to make the paper self-contained.

We now turn to a description of the HMC algorithm itself. For brevity, we will only give a summary of the essential steps here and refer the reader interested in a step-by-step derivation to Ref.~\cite{Smekal:1403.3620}. In HMC, the generation of representative configurations of the  $\phi$ field consists of two parts: The first is a molecular dynamics (MD) trajectory in which $\phi$ is evolved in computer time through an artificial Hamiltonian dynamics. To this end, a conjugate momentum $\pi$ is introduced which is refreshed with Gaussian noise at the beginning of each trajectory, and the classical Hamilton equations for $\phi$ and $\pi$ are integrated using a symplectic integrator. Since this introduces a numerical error associated with finite integration steps, a Metropolis accept/reject step is then used to make the algorithm exact. 

Typically, the fermion determinant is sampled stochastically using pseudofermions, both for force calculations during the MD trajectories and for calculations of the total action during the Metropolis step. The bulk of the results in this work were obtained using this method. Another order of magnitude increase in performance is possible in principle by avoiding the use pseudofermions altogether and using exact derivatives of the fermion determinant instead. A small fraction of our results was obtained using this technique, but this is a very recent development and will be described in a separate publication. 

Lastly, we would like to point out that HMC simulations using a single Hubbard field can suffer from a loss of ergodicity if no additional mass terms are included in $\hat{\mathcal{H}}$. The reason is the presence of extended manifolds with $\det M(\phi)=0$ in configuration space, which form barriers separating regions of $\textrm{sgn}(\det M(\phi))=\pm 1$ and which exhibit divergences in the effective potential through which the molecular dynamics cannot tunnel, except on small lattices. That this is a problem in practice, in particular at low temperatures, was shown in Refs.
\cite{Ulybyshev:1712.02188,Assaad:1708.03661}.\footnote{We remark here that with the linearized fermion action the leading discretization errors mimic the effect of a mass term and thus restore ergodicity at finite $\delta_\tau$. This feature is not useful in practice however, since ergodicity is nevertheless lost as the continuum limit is approached and potential barriers become a problem precisely when $N_\tau$ is sufficiently large for the linearized action to be reliable.}

One way to avoid this problem is to extend the configuration space to complex numbers. This does not remove the barriers, but the additional degrees of freedom allow MD trajectories to circumvent them. To achieve this we rewrite the on-site interaction term as
\begin{eqnarray}
\label{eq:complexint}
\frac{U}{2} \hat{\rho}_x^2 =
\alpha \frac{U}{2} \hat{\rho}_x^2 - (1-\alpha) \frac{U}{2}(\hat{\rho}_x')^2 + U(1-\alpha)\,\hat{\rho}_x',
\end{eqnarray} where $\hat{\rho}_x' = \hat{a}^{\dagger}_x \hat{a}_x + \hat{b}^{\dagger}_x \hat{b}_x$ is the spin-density operator. Consider now, that an equally valid variant of the HS transformation is
\begin{eqnarray}
 \label{continuous_HS_real}
   e^{\frac{\delta_\tau}{2}\sum_{x,y} V_{x,y} \hat \rho_x \hat \rho_y} \cong \int D \chi\, e^{- \frac{1}{2\delta_\tau} \underset{x,y}{\sum} \chi_x V^{-1}_{xy} \chi_y} e^{ \underset{x}{\sum} \chi_x \hat \rho_x},
\end{eqnarray}
where in contrast to Eq.~(\ref{continuous_HS_imag}) the last exponent is purely real. By applying 
Eq.~(\ref{continuous_HS_imag}) to the first term and Eq.~(\ref{continuous_HS_real}) to the second term we obtain a Hubbard field which has real and imaginary components. By choosing $\alpha \in [0,1]$ we can interpolate between the purely real and purely imaginary cases. 

The exponents $e^{i \phi_{x,t}}$ in the  fermion matrix (\ref{eq:fermionmatrix}) are now replaced by $e^{i \phi_{x,t} + \chi_{x,t}}$, and the contribution of the on-site interaction term to the action of Hubbard-Stratonovich fields becomes
\begin{eqnarray}
   S_\alpha(\phi,\chi) = \sum_{x,t}\left(  \frac  {\phi_{x,t}^2} {2 \alpha \delta_\tau U}  + \frac {(\chi_{x,t}- (1-\alpha) \delta_\tau U)^2} {2 (1-\alpha) \delta_\tau U} \right).
  \label{eq:action_alpha}
\end{eqnarray}
The constant shift of $\chi$ results from the last term in Eq.~(\ref{eq:complexint}).
Note that applying the above procedure to the on-site potential only, without changing the treatment of the non-local parts of the interaction potential $V_{xy}$, is entirely sufficient to obtain an ergodic algorithm. Also note that the introduction of the complex fields changes the restriction on the interactions to $V<\alpha U/3$.

A substantial part of the results in this work were obtained using $\alpha=1$. We do however carry out extensive cross-checks using the two-field formalism in order to rule out a strong effect of ergodicity problems on our results, and quantify the systematic errors where appropriate. The main insight is that in our study of spin-density wave order, the qualitative picture remains entirely intact and critical exponents and the location of the phase boundary are only very weakly affected. On the other hand, in the one-field formalism there is a tendency towards an unphysical charge ordering. We discuss these issues in detail in Section~\ref{sec:Results} and explain how we have verified that CDW order is indeed absent. 

All results in this work were obtained at temperatures $T=0.125\textrm{eV}=0.046 \kappa$ with $N_\tau=128$, which leads to a time discretization $\delta_\tau=0.16 \kappa^{-1}$. Previous experience \cite{Buividovich:16:4} has shown this to be sufficient to strongly suppress discretization errors when using the fully exponential fermion matrix (\ref{eq:fermionmatrix}) with exact particle-hole symmetry and hence without spin-symmetry violations. For each lattice configuration we compute the full fermionic equal-time Green function $g(x,y)=\langle\hat{a}_x \hat{a}^{\dagger}_y\rangle=M^{-1}_{x,t,y,t}$ and then express other observables in terms of these (see Appendix \ref{apdx:Observables} for explicit expressions). To account for possible autocorrelation effects in our data, we use binning to calculate statistical errors. Typical sample sizes are on the order of several hundreds of independent measurements. 

\section{Results}
 \label{sec:Results}

\subsection{Spin-density wave order}
 \label{subsec:sdw}

To detect ordered phases we employ two distinct methods: extrapolating an order parameter to the thermodynamic limit, and analyzing its finite-size scaling in the vicinity of the presumed phase boundary, as estimated  using  the  first  method. By demonstrating consistency between these two approaches we can establish the existence of an ordered phase in the $U-V$ plane with high confidence, determine its boundary and study the critical properties thereof.

The SDW phase is characterized by separation of electron spins between the two sublattices, with the difference of spins between the sublattices being the order parameter. This order parameter, however, vanishes in a finite volume, and can only be recovered by introducing a small ``seed'' perturbation, which favours spontaneous symmetry breaking in this specific direction and which must then be taken to zero while extrapolating the order parameter to the thermodynamic limit. While this method was used in our previous HMC simulations \cite{Buividovich:12:1,Smekal:1403.3620,Buividovich:13:5}, in this paper we avoid such an approach. To carry out an unbiased study of competing ordered phases, the use of such perturbations is unfeasible for a number of reasons: First and foremost, each choice of source term leads to a bias towards a particular phase and does not allow for the detection of other phases. This implies that the required extrapolations, which are computationally very expensive as different lattice sizes must be simulated for several different values of the external source, must be repeated for each of the different phases under investigation. Furthermore, the extrapolations themselves can also carry some ambiguity as the exact scaling-laws with which the combined zero-source and thermodynamic limits are approached are typically non-linear and not known. And finally, the implementation of such sources in the HMC simulation is not always straight-forward and in some important cases, such as a CDW phase in the Hubbard model, even leads to a fermion sign-problem which prevents the use of HMC altogether.\footnote{CDW order is induced by a sublattice-staggered mass term of the form $\sum_x m_s (\hat{a}^{\dagger}_x \hat{a}_x - \hat{b}^{\dagger}_x \hat{b}_x)$ in the notation of Section \ref{sec:Setup}, where the sign of $m_s$ alternates between the sublattices. Due to the relative minus sign between $\hat{a}^{\dagger}_x \hat{a}_x$ and 
$\hat{b}^{\dagger}_x \hat{b}_x$ the fermion matrices for spin-up and spin-down electrons are no longer Hermitian-conjugate pairs when including such a source.}  

Instead, in this work we infer the phase structure from the volume dependence of quadratic observables which are non-zero in finite volume even without external sources. To detect SDW, we use the square of the total spin per sublattice
\begin{eqnarray}
\label{eq:squarespin}
 \langle S_i^2 \rangle  
 =  
 \left\langle \frac{1}{L^4} \left( \sum\limits_{x\in A} \hat{S}_{x,i} \right)^2 \right\rangle 
 + \nonumber \\ +
 \left\langle \frac{1}{L^4} \left( \sum\limits_{x\in B} \hat{S}_{x,i} \right)^2 \right\rangle , 
\end{eqnarray}
where $L$ is the linear lattice size and
\begin{eqnarray}
 \hat{S}_{x,i} = \frac{1}{2} ( \hat
 c^\dag_{x, \uparrow} , \, \hat c^\dag_{x, \downarrow} )  \sigma_i
 \left(
 \begin{array}{c}
  \hat{c}_{x, \uparrow} \\ 
  \hat{c}_{x, \downarrow} \\
 \end{array} 
 \right)
\end{eqnarray}
is the $i$-th component of the spin operator at lattice site $x$. Due to the exact spin-symmetry of (\ref{eq:fermionmatrix}), the choice of $i$ is irrelevant as was explicitly verified in Ref. \cite{Buividovich:16:4}. In (\ref{eq:squarespin}) we have also used the equivalence between the two sublattices $A$ and $B$ and added the corresponding observables together, which improves the signal-to-noise ratio in Monte-Carlo simulations. An explicit expression for the expectation value (\ref{eq:squarespin}) in terms of fermionic Green functions is given in Appendix \ref{apdx:Observables}.

\begin{figure*}[h!tpb]
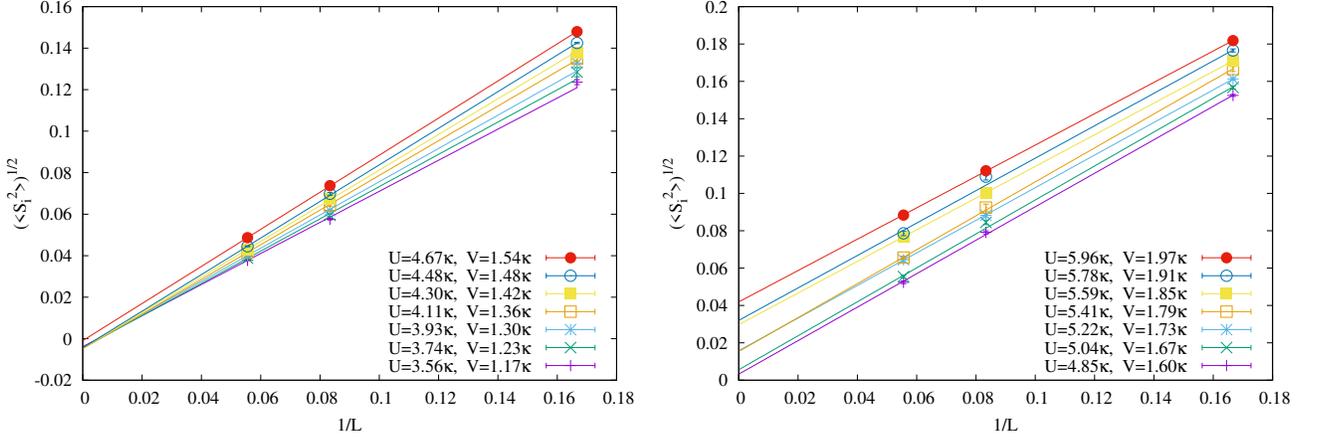

  \includegraphics[width=0.48\textwidth]{{{extrapolation_plots_diagonal_full_spin_1}}}
  \includegraphics[width=0.48\textwidth]{{{extrapolation_plots_diagonal_full_spin_2}}}\\
    \caption{ Linear $L \to \infty$ extrapolation of $\sqrt{\langle S_i^2 \rangle}$ for $(U,V)$ values along the $V=U/3$ line. On the left: in the weak-coupling regime, on the right: in the strong-coupling regime with SDW order.}
    \label{fig:extrapolation_Sx_1}
\end{figure*}

\begin{figure*}[h!tpb]
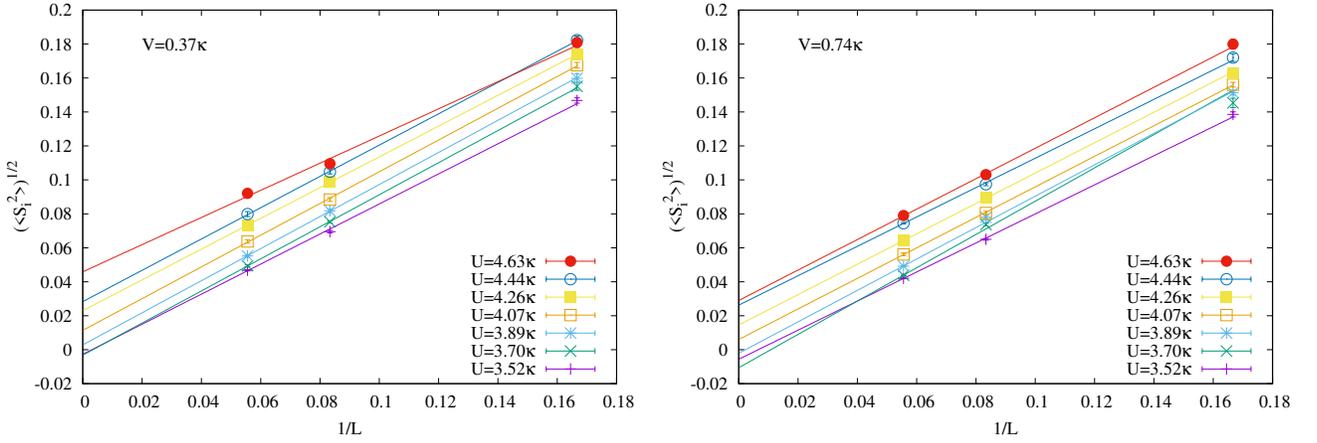

  \includegraphics[width=0.48\textwidth]{{{extrapolation_plots_V01_1.000000e+00_full_spin_1}}}
  \includegraphics[width=0.48\textwidth]{{{extrapolation_plots_V01_2.000000e+00_full_spin_1}}}\\
    \caption{Linear $L \to \infty$ extrapolation of $\sqrt{\langle S_i^2 \rangle}$ for $V=0.37 \kappa$ (left) and $V=0.74 \kappa$ (right).}
    \label{fig:extrapolation_Sx_2}
\end{figure*}

In order to detect the ordered phase, we first consider the infinite-volume extrapolations of the quantity $\sqrt{\langle S_i^2 \rangle}$. In the phase with an antiferromagnetic ordering it should extrapolate to a finite value, and otherwise it should extrapolate to zero. This extrapolation procedure is similar in spirit to the one used in \cite{Assaad:1304.6340}. Away from a phase transition $\sqrt{\langle S_i^2 \rangle}$ is expected to depend on the lattice size as $\sqrt{\langle S_i^2 \rangle} = a L^{-1} + b$. In principle the leading power of $L$ in this expression should deviate from $L^{-1}$ close to the phase boundary, where it is replaced by a critical finite-size scaling relation, but we nevertheless find that linear fits using $L=6,12,18$ work well for all points in the $U-V$ plane considered. The linear fit was also verified for several points using additional lattice sizes ($L=8,14$). In this case higher than linear powers can be included into the fitting function, but it appears that they do not add to the goodness of the fit. This already hints that the exact critical exponent cannot be too far from unity. 

\begin{figure*}[h!tpb]
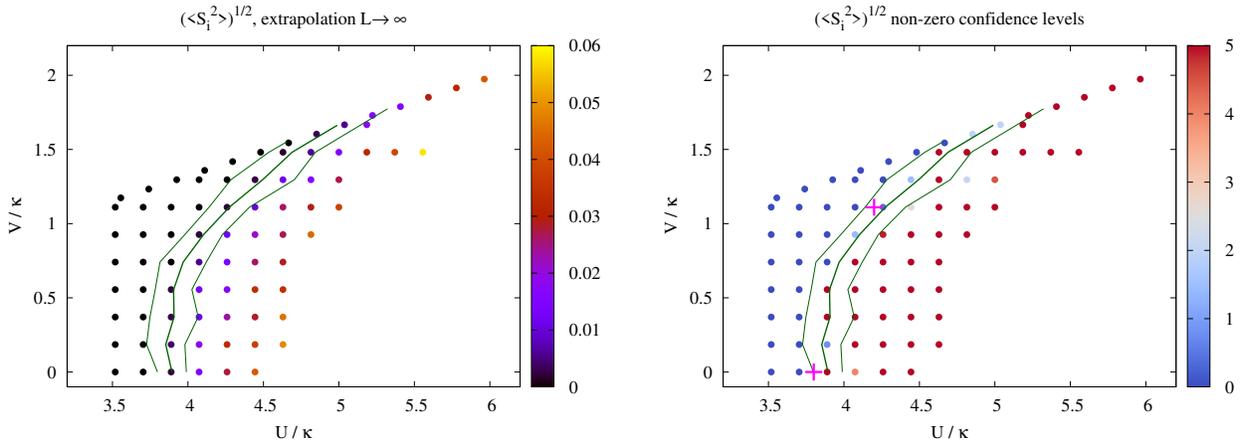

  \includegraphics[width=0.48\textwidth]{{{Sx_inf_2x128_withboundary_T0.0625_new}}}
  \includegraphics[width=0.48\textwidth]{{{Sx_phasediag_inf_2x128_withboundary_T0.0625_new}}}\\
    \caption{On the left: $L\to \infty$ limit of $\sqrt{\langle S_i^2 \rangle}$ from $L=6,12,18$ (dots). On the right: number of standard deviations with which $\sqrt{\langle S_i^2 \rangle}$ is non-zero at $L= \infty$ (values $>5$ are displayed with the same color as $5$). On both plots we also show the phase boundary from intersection method with $2 \sigma$ confidence band (lines). Crosses mark $U_c/\kappa=3.78$  (prediction of Ref. \cite{Assaad:1304.6340}, bottom cross) and result from simulation with complex Hubbard field ($\alpha=0.95$, top cross).}
    \label{fig:phase_diag_Sx}
\end{figure*}

We carry out the $L \to \infty$ extrapolation using the fits of the form $f(1/L)=a L^{-1} + b$ with lattice sizes $L=6,12,18$ for a large set of points in $U-V$ space, using HMC data obtained with a single Hubbard-Stratonovich field ($\alpha=1$ in the notation of Section~\ref{sec:Setup}). Figs.~\ref{fig:extrapolation_Sx_1} and \ref{fig:extrapolation_Sx_2} show such extrapolations for several points on the $U-V$ phase diagram along the line $V=U/3$ (simulations exactly on this line are not possible so all points are shifted slightly away from this line) and along two $V= \const$ lines, respectively. Fig.~\ref{fig:phase_diag_Sx} (left) summarizes the results of such $L \to \infty$ extrapolations for all values of $U$ and $V$ which we have considered. 

To establish the ordered phase quantitatively, we use the statistical error of the constant $b$ as obtained from the fit. Fig. \ref{fig:phase_diag_Sx} (right) shows the number of standard deviations with which a non-zero value of $\sqrt{\langle S_i^2 \rangle}$ is obtained for each point. We find SDW order at $>5\sigma$ confidence at sufficiently large $U$ for all $V$ values considered, with a rather sharp boundary which curves towards larger values of $U$ when $V$ is increased. Within our resolution the $V=0$ results are consistent with the value $U_c/\kappa=3.78$ obtained in \cite{Assaad:1304.6340}.

We note here in passing that
$\sqrt{\langle S_i^2 \rangle}>0$ can in principle also indicate a ferromagnetic phase. To uniquely identify SDW order, we also measure the mean squared magnetization
\begin{eqnarray}
\label{eq:ferro}
 \langle m_i^2 \rangle  
 =  
 \left\langle \frac{1}{L^4} \left( \sum\limits_{x} \hat{S}_{x,i} \right)^2 \right\rangle, 
\end{eqnarray}
for each parameter set (for an expression of $\langle m_i^2 \rangle$ in terms of Green functions see Appendix \ref{apdx:Observables}). We find that $\sqrt{\langle m_i^2 \rangle}$ is at least an order of magnitude smaller than $\sqrt{\langle S_i^2 \rangle}$ for each point in the $U-V$ plane considered and each lattice size $L$ (this has been verified both for $\alpha=1.0$ and the case $\alpha=0.95$ discussed further below). Moreover, linear $L\to\infty$ extrapolations of $\langle m_i^2 \rangle$ yield results consistent with zero in all cases. See Fig. \ref{fig:extrapolation_ferro} for examples.

\begin{figure}[h!tpb]
  \includegraphics[width=0.48\textwidth]{{{extrapolation_plots_V01_5.000000e-01_full_ferro_1}}}\\
  \caption{
   Linear $L \to \infty$ extrapolation of $\sqrt{\langle m_i^2 \rangle}$ for $V=0.19 \kappa$ with $\alpha=1.0$.
  }
    \label{fig:extrapolation_ferro}
\end{figure}

While infinite-volume extrapolation detects the ordered phase, it cannot distinguish a disordered phase from a region with large statistical errors. Furthermore, the extrapolation does not tell us anything about the nature of the phase boundary. In order to complement our extrapolation analysis, we also study the finite-size scaling of the squared spin per sublattice (\ref{eq:squarespin}). Ref. \cite{Assaad:1304.6340} verified the finite-size scaling law $m = L^{-\beta/\nu}F(L^{1/\nu}(U-U_c))$ for the staggered magnetization at $V = 0$ and obtained $\beta/\nu \simeq 0.89$, in agreement with the chiral Heisenberg Gross-Neveu universality class. The corresponding scaling law for $\langle S_i^2 \rangle$ at $U \equiv U_c$ is  $\langle S_i^2 \rangle = c \, L^{-2\beta/\nu}$. With properly chosen  $\beta/\nu$, we should be able to exactly obtain the phase boundary in the entire $U-V$ plane by locating intersection points of the functions $\langle S_i^2 \rangle L^{2\beta/\nu}$ for different $L$ when traversing the $U-V$ plane along different lines.

It is a priori not clear that the same $\beta/\nu$ applies at each point of the phase boundary. What is needed is an unbiased method to determine both $\beta/\nu$ and the intersection points from the data, preferably with estimates of the statistical error. We describe such a method in the following. 

To carry out a proper scaling analysis, we first note that the data points in Fig.~\ref{fig:phase_diag_Sx} (right) show a rather sharply bound region of non-zero $\langle S_i^2 \rangle$. Thus we have probable cause to expect a scaling window in the border region. For a given line in the $U-V$ plane we now identify a region around the presumed boundary in which $\langle S_i^2 \rangle L^{2\beta/\nu}$ has an approximately linear dependence on the external parameter ($U$, $V$ or a combination thereof) for all $L$. This is done by manual tuning of the window. To estimate $\beta/\nu$, we then use linear fits to the data of the form $\langle S_i^2 \rangle L^{2\beta/\nu}= a x + b$ (where $x$ denotes a generic external parameter) and adjust $\beta/\nu$ until the enclosed triangle between the lines modelling the $L=6,12,18$ data is minimized. Furthermore, the upper and lower bounds of the fit windows are also varied independently until an optimal intersection is obtained. For each of our data sets we find that some choice of $\beta/\nu$ and fit window yields an unambiguous optimum.

\begin{figure*}[h!tpb]
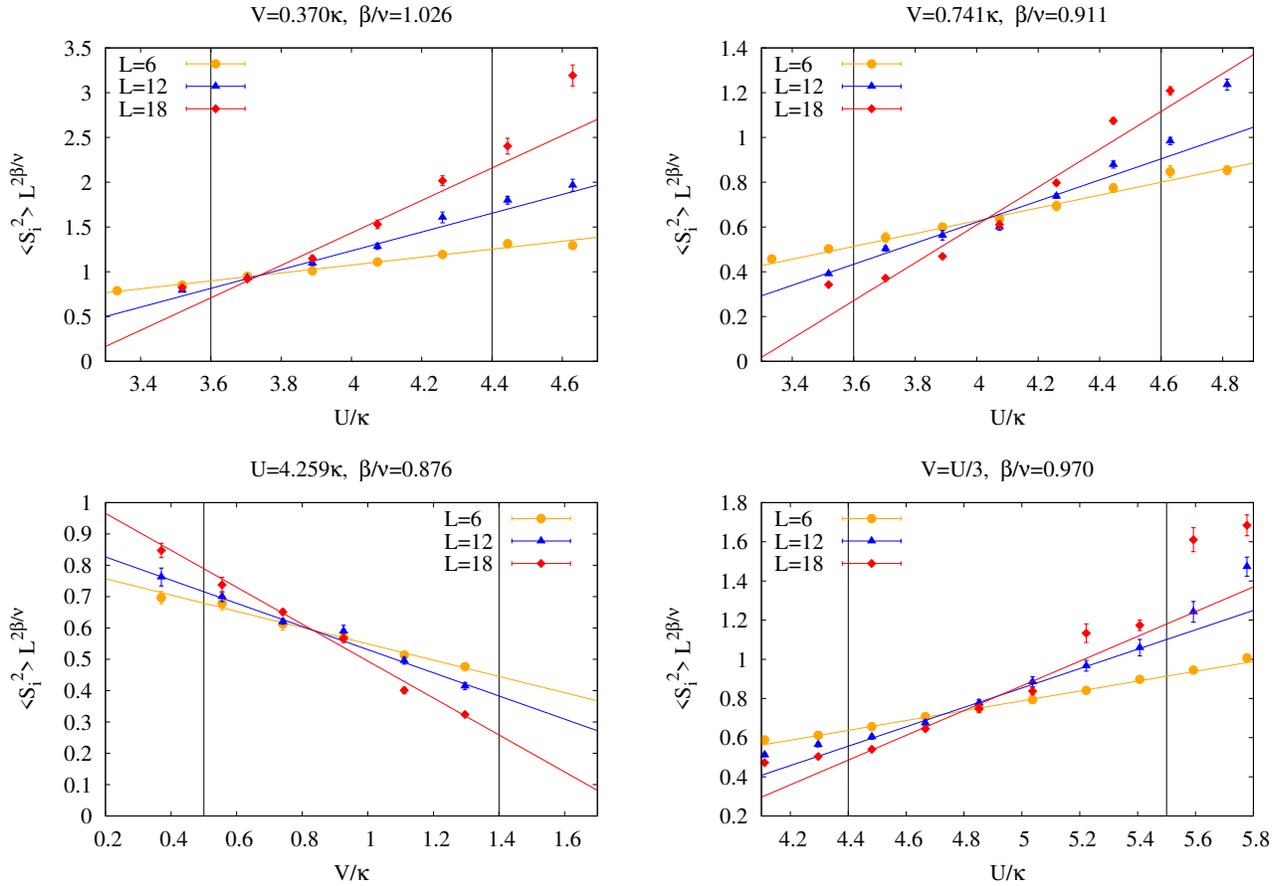

  \includegraphics[width=0.48\textwidth]{{{Sx_scaling_T0.0625_V01_1.000000e+00_optimal_withfit_zeromodes}}}
  \includegraphics[width=0.48\textwidth]{{{Sx_scaling_T0.0625_V01_2.000000e+00_optimal_withfit_zeromodes}}}\\
  \includegraphics[width=0.48\textwidth]{{{Sx_scaling_T0.0625_V00_1.150000e+01_optimal_withfit_zeromodes}}}
  \includegraphics[width=0.48\textwidth]{{{Sx_scaling_T0.0625_diagonal_optimal_withfit_zeromodes}}}\\
    \caption{Optimized intersection of $\langle S_i^2 \rangle L^{2\beta/\nu}$ with $L=6,12,18$
        for $V=0.37 \kappa$ (top left), $V=0.74 \kappa$ (top right), $U=4.26 \kappa$ (bottom left)  and $V\approx U/3$ (bottom right). Vertical lines mark the windows in which linear fits were applied.}
    \label{fig:scaling_Sx2}
\end{figure*}

We apply this procedure to the full set of horizontal ($V = \const$) lines in the $U-V$ plane up to $V=1.48\kappa$, as well as along the $V=U/3$ line and the vertical lines $U=4.07\kappa, 4.25\kappa, 4.44\kappa$. We find that the procedure works well for all sets of data points considered (the enclosed triangles are very small in all cases and the intersection points all fall in the immediate vicinity of the presumed boundary), as illustrated on Fig.~\ref{fig:scaling_Sx2} for several characteristic points in the $U-V$ phase diagram.

\begin{figure}[h!tpb]
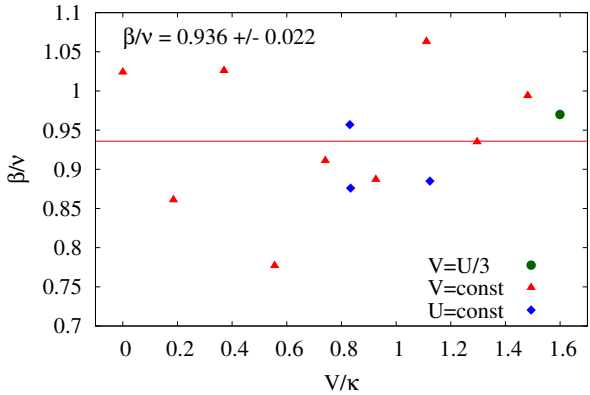

  \includegraphics[width=0.48\textwidth]{{{Sx_scaling_exponents_zeromodes}}}\\
    \caption{Critical exponents from intersection analysis along presumed phase boundary. For $U=const$ and $V\approx U/3$ lines the $V_c$ estimate is chosen as the $x$-value.}
    \label{fig:expoents_Sx2}
\end{figure}

Since all data sets are affected by statistical errors, the optimized $\beta/\nu$ can be interpreted as random variables, drawn from some probability distribution around the true value.\footnote{In principle there is also a systematic uncertainty associated with the choice of scaling window. By allowing a variation of the bounds of the window during our optimization procedure we have traded this for an additional statistical error.} To get a sense of how $\beta/\nu$ depends on the location in the $U-V$ plane, we track how the optimized values change along the presumed phase boundary. Fig.~\ref{fig:expoents_Sx2} shows a collection of $\beta/\nu$ values obtained along the lines $V = \const$, $U = \const$ and $V = U/3$. From left to right plots, these values are traced along the boundary from the $V=0$ to the $V=U/3$ line. What we find, is the absence of any noticeable trend: Our $\beta/\nu$ estimates all appear to be distributed around some mean value. This strongly suggests that the entire phase boundary is characterized by the same critical behavior. Under the assumption that the same critical exponent applies everywhere, we can consider each data point as an independent measurement (as separate data sets were used in each case) and estimate $\beta/\nu=0.936\pm 0.022$. The value $\beta/\nu=0.89$, obtained in Ref.~\cite{Assaad:1304.6340}, is  $\sim 5 \%$ smaller and falls right onto our lower $2\sigma$ limit. Much larger lattices and sample sizes would be needed to clearly decide whether this small discrepancy is a statistical fluctuation, a finite-size effect, or a consequence of the ergodicity violation described in Section~\ref{sec:Setup}. We note that our errorbar only accounts for the statistical uncertainty of the optimization procedure and not for the (probably larger) systematic uncertainty of the limited lattice sizes.

\begin{table}[h]
\centering
\begin{tabular}{c|l|l|l}
\hline\hline  \rule{0pt}{2.6ex}               
 & $~~1/\nu~~$ & $~~\beta/\nu~~$  & $~~\nu~~$  \\[0.5ex]
\hline  \rule{0pt}{2.6ex}          
$\epsilon$ expansion [2,2] Pad\'e \cite{HerbutScherer2017} & $0.6426$ & $0.99925$ &   \\
$~\epsilon$ expansion [3,1] Pad\'e \cite{HerbutScherer2017} & $0.6447$ & $0.97815$ &  $1.2352$ \\
 Functional RG \cite{Knorr:2018} & $0.795$ & $1.016$ &  $1.26$\\
  Large N \cite{Gracey:1801.01320}  & $0.8458$ & $1.09245$ &  $1.1823$\\[1ex] 
  \hline \rule{0pt}{2.6ex}          
  Monte-Carlo \cite{PhysRevX.6.011029} &  & $0.74(2)$ &  $1.02(1)$\\
 $~$Monte-Carlo \cite{PhysRevB.91.165108}  &  & $0.85(8)$ &  $0.84(4)$\\[1ex] 
\hline\hline
\end{tabular}
\caption{
Adapted from \cite{Gracey:1801.01320}: Critical exponents of the continuous $N=2$ chiral Heisenberg Gross-Neveu model
in three spacetime dimensions obtained from renormalization group studies (top) and of related discrete Hubbard-type models obtained from Monte-Carlo simulations (bottom). Where $\nu$ and $1/\nu$ are both displayed they were determined independently.}
\label{tab:exponents}
\end{table}

We point out here that critical exponents for the universality class of the $N=2$ chiral Heisenberg Gross-Neveu field-theory in three spacetime dimensions which presumably applies to this antiferromagnetic phase transition are not known to great numerical precision. Latest results from $1/N$ expansion \cite{Gracey:1801.01320}, functional renormalization group \cite{Knorr:2018}  and $\epsilon$-expansion \cite{HerbutScherer2017} in aggregate suggest roughly $\beta/\nu\approx 1$ (see Table \ref{tab:exponents} for summary). Our result is slightly smaller but likely falls within the bounds of theoretical uncertainty (our upper $2\sigma$ limit of $\beta/\nu=0.98$ certainly does). Also, slightly smaller values tend to be observed in Monte-Carlo simulations of related discrete Hubbard-type models believed to fall into this universality class \cite{PhysRevX.6.011029,PhysRevB.91.165108,PhysRevB.90.085146}. To obtain additional evidence that we are indeed seeing the critical behavior of this second-order transition we also verify the corresponding collapse of the data on a universal finite-size scaling function $f(x)$,
\begin{equation}
\langle S_i^2 \rangle = L^{-2\beta/\nu}  f(L^{1/\nu}\epsilon)\, ,
\end{equation}
where $\epsilon $ is the reduced coupling used as the control parameter, and extract the correlation-length exponent $\nu$, for which the methods cited above, on average, suggest $\nu\approx 1.2$.

\begin{figure}[h!tpb]
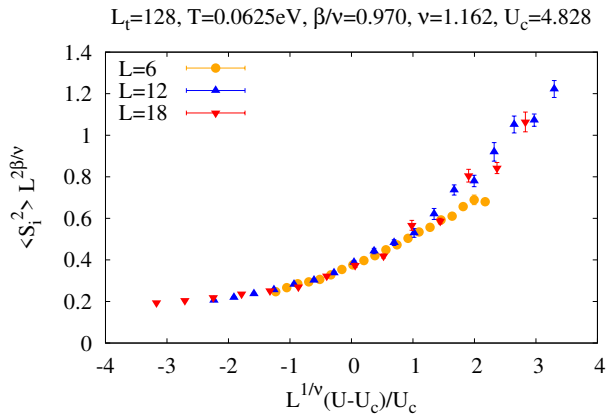

  \includegraphics[width=0.48\textwidth]{{{Sx_collapse_T0.0625_diagonal_optimal_withfit_zeromodes}}}
    \caption{Critical scaling of $\langle S_i^2 \rangle$ at $\alpha=1$ along the $V=U/3$ line. $\nu$ and $U_c$ are obtained by optimizing the $\chi^2/\textrm{dof}$ of
    a polynomial fit to data from all lattice sizes within the scaling window  shown in Fig. \ref{fig:scaling_Sx2}  }
    \label{fig:spin_collapse}
  \end{figure}

Fig.~\ref{fig:spin_collapse} shows an optimized collapse where we fit data points from $L=6,12,18$ with a polynomial function of $x=L^{1/\nu}(U-U_c)/U_c$ and adjust both $U_c$ and $\nu$ until the $\chi^2$ per degree of freedom becomes minimal. This is illustrated  here for the data along the $V = U/3$ line shown in the bottom right panel of Fig.~\ref{fig:scaling_Sx2}, where we have the largest statistics. We choose the same scaling window as in Fig.~\ref{fig:scaling_Sx2}, use the same value $\beta/\nu=0.97$ that results from the intersection method for this line, and then obtain $\nu=1.162$ which is inline with the theoretical predictions. As a consistency check, the resulting $U_c=4.828$ is in good agreement with that obtained from the intersection method in Fig.~\ref{fig:scaling_Sx2}, as discussed in the following paragraph. The deviations from finite-size scaling, observed in Fig.~\ref{fig:spin_collapse} above $x \approx 1 $, are typical of the expected corrections to scaling at small $L$ as well.     

We therefore conclude with some confidence that what we are seeing is at least consistent with critical scaling in the chiral Heisenberg Gross-Neveu universality class. The same conclusion, with somewhat larger uncertainties but no systematic deviations, is also obtained for the other data sets of Fig.~\ref{fig:scaling_Sx2}. We certainly observe no significant changes in the scaling behavior along the whole transition line shown in Fig.~\ref{fig:phase_diag_Sx}, starting from $V=0$ to the $V=U/3$ line used as our representative example in Fig.~\ref{fig:spin_collapse}.

 Finally, let us determine the phase boundary from the intersection points of the linear fits of the data for $\langle S_i^2 \rangle L^{2\beta/\nu}$ and estimate the corresponding error band. Instead of using the individual values obtained from the optimization, we do the following: For each horizontal line and for $V=U/3$ 
 we set $\beta/\nu$ to $0.958$ and subsequently to $0.914$, which corresponds to our upper and lower one-$\sigma$ limits respectively. For each choice, we obtain the intersection points of $L=\{6,12\}$, $L=\{6,18\}$ and $L=\{12,18\}$. This gives $6$ estimates for position of the phase boundary along this line. Of these we use the sample mean as our final answer and the standard deviation of the sample to quantify the statistical uncertainty (we do not use the standard error of the mean here, as the same raw data are re-used to obtain multiple estimates of $U_c$). By repeating this for every line, we obtain a phase boundary together with a confidence band, which is shown in Fig.~\ref{fig:phase_diag_Sx} together with the results of the extrapolation of $\sqrt{\langle S_i^2 \rangle}$. We find a striking coincidence between the two methods which lends solid credibility to our results. 
 
In particular, for $V=0$ we find $U_c/\kappa=3.9\pm 0.05$. The value $U_c/\kappa=3.78$ obtained in Ref.~\cite{Assaad:1304.6340}, and marked by a cross in Fig.~\ref{fig:phase_diag_Sx}, differs by $\sim 3\%$ and falls just outside of our lower $2\sigma$ limit. This small difference is likely due to ergodicity violations in our massless simulations with a single Hubbard field $\phi$ corresponding to $\alpha = 1$ in Sec.~\ref{sec:Setup}. The magnitude of the discrepancy is consistent with the results of Ref.~\cite{Ulybyshev:1712.02188} where it was shown that $\langle S_i^2 \rangle$ changes only by a few percent close to $U_c$ at $V=0$ if one shifts the mixing parameter $\alpha$ in the range $[1,0.9]$. 

 \begin{figure*}[h!tpb]
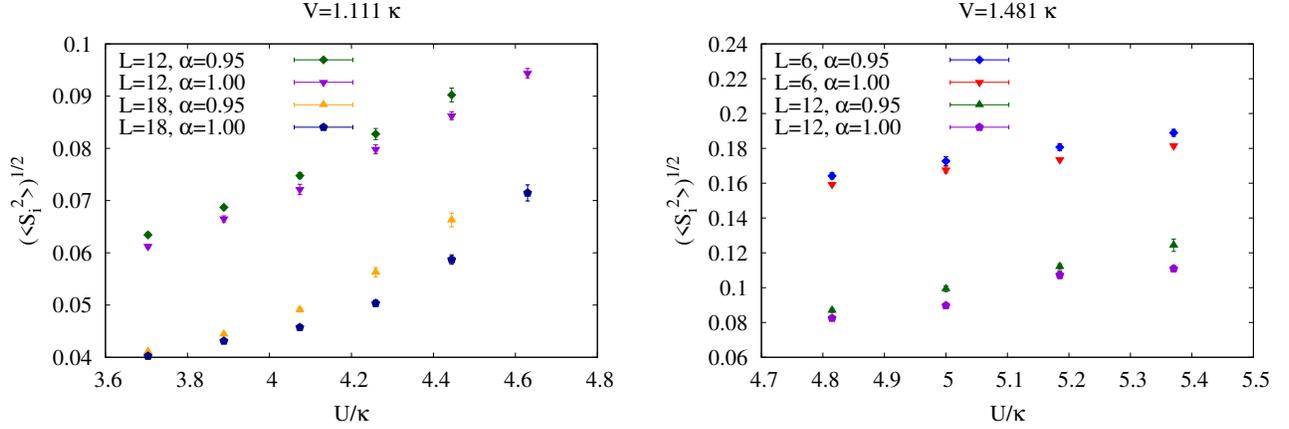

  \includegraphics[width=0.48\textwidth]{{{Sx_twofield_comparison_L_12_18_V3.00}}}
  \includegraphics[width=0.48\textwidth]{{{Sx_twofield_comparison_L_6_12_V4.00}}}\\
 \caption{Comparison of $\sqrt{\langle S_i^2 \rangle}$ as obtained from simulations with one ($\alpha=1.0$) and two ($\alpha=0.95$) Hubbard fields.}
    \label{fig:Sx_twofield_comparison}
\end{figure*}

We therefore now verify that non-ergodicity of our simulations affects the results for $\langle S_i^2 \rangle$ at $V \neq 0$ not any stronger than at $V = 0$. 
 To this end, we first determine which choice of $\alpha$ can be considered safe for ergodic simulations. In Ref.~\cite{Ulybyshev:1712.02188} it was shown that simulations at $V = 0$ are essentially ergodic for $\alpha \lesssim 0.95$ with $L=6$, but it is unclear
whether this carries over to $V \neq 0$ and larger lattices. To clarify this we carry out simulations on $L=12$ lattices for $3$ points close to the $V=U/3$ line with $\alpha=[0.925,0.99]$. We choose $(U/\kappa,V/\kappa)$ as $(3.70,1.11)$, 
$(4.44,1.29)$ and $(5.37,1.48)$ which fall deeply in the disordered phase, close to the presumed phase boundary and deeply in the ordered phase respectively. In each case we compute $\langle S_i^2 \rangle$ and $\langle q^2 \rangle$ (introduced in Subsection~\ref{subsec:cdw}) and find no statistically significant dependence on $\alpha$ for either observable. We thus conclude that the safe range extends to even larger $\alpha$ than for the case $V=0,L=6$. 

\begin{figure}[h!tpb]
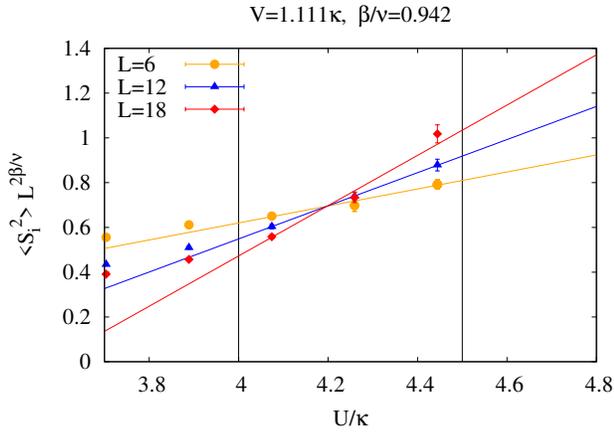

  \includegraphics[width=0.48\textwidth]{{{Sx_scaling_T0.0625_V01_3.000000e+00_optimal_withfit_zeromodes_twofields}}}
    \caption{Results with complexified Hubbard field ($\alpha=0.95$): Optimized intersection of $\langle S_i^2 \rangle L^{2\beta/\nu}$ with $L=6,12,18$ at $V=1.111\kappa$.}
\label{fig:twofields_scaling_V=1.111}
  \end{figure}

Fig. \ref{fig:Sx_twofield_comparison}
shows a direct comparison of the data obtained in the one-field formalism and a new set of data, subsequently obtained with $\alpha=0.95$. The figures show the $U$ dependence of $\sqrt{\langle S_i^2 \rangle}$ for the lines $V = 1.111\kappa$ and $V = 1.481\kappa$ with different lattice sizes. We observe that the inflection points (corresponding approximately to $U_c$) shift at most by a few percent when introducing the complex Hubbard field. We then repeat the finite-size scaling analysis
of $\langle S_i^2 \rangle$ 
for the $V = 1.111\kappa$ line with $\alpha=0.95$, using lattice sizes $L=6,12,18$. Fig.~\ref{fig:twofields_scaling_V=1.111} shows the result of the area minimization procedure (as described above) for this case.  The critical exponent evaluates to $\beta/\nu=0.942$ which falls within one standard deviation of our estimate using the single Hubbard field. We find $U_c/\kappa=4.20$ which falls barely above the lower $2\sigma$ limit
 of our phase boundary (see Fig.~\ref{fig:phase_diag_Sx} where this point is marked by the second cross). We thus conclude that observables characterizing the SDW order are indeed only weakly affected by the non-ergodicity of the standard HMC algorithm in the massless limit, similar to the case $V=0$.
 
\subsection{Charge-density wave order}
 \label{subsec:cdw}

To study CDW order we define the squared charge per sublattice as
\begin{eqnarray}
\label{eq:squarecharge}
 \langle q^2 \rangle  
 =  
 \left\langle \frac{1}{L^4} \left( \sum\limits_{x \in A} \hat{\rho}_x \right)^2 \right\rangle
 +
 \left\langle \frac{1}{L^4} \left( \sum\limits_{x \in B} \hat{\rho}_x \right)^2 \right\rangle , 
\end{eqnarray}
in full analogy with the definition (\ref{eq:squarespin}) of the squared spin $\langle S_i^2 \rangle$. As in the previous Subsection~\ref{subsec:sdw}, we use this observable, again expressed in terms of fermionic Green functions in Appendix \ref{apdx:Observables}, now to detect possible CDW order by combining the $L \to \infty$ extrapolation of $\sqrt{\langle q^2 \rangle}$ and the finite-size scaling analysis of $\langle q^2 \rangle$. 

\begin{figure}[h!tpb]
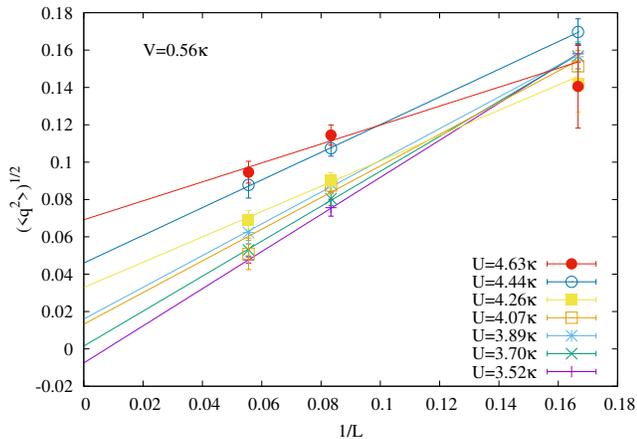

  \includegraphics[width=0.48\textwidth]{{{extrapolation_plots_V01_1.500000e+00_full_charge_1}}}\\
  \caption{
    Linear $L \to \infty$ extrapolation of $\sqrt{\langle q^2 \rangle}$ for $V=0.56 \kappa$ with $\alpha=1.0$.}
    \label{fig:extrapolation_q}
\end{figure}

Our first observation in simulations with one Hubbard field ($\alpha=1$) is that the statistical error of charge observables is much larger than that of spin observables. This already foreshadows problems. We nevertheless are able to carry out the $L \to \infty$ extrapolations and apply the intersection method, finding that CDW in general seems to coincide with the existence of SDW order (to give one example, Fig. \ref{fig:extrapolation_q} shows how for $V=0.56 \kappa$ the extrapolated $\sqrt{\langle q^2 \rangle}$ becomes non-zero at $U\gtrsim 4.0 \kappa$). We obtain a phase diagram for CDW that looks very similar to Fig.~\ref{fig:phase_diag_Sx}, but with much more noise along the presumed phase boundary. The critical exponent obtained from the intersection method evaluates to $\beta/\nu \approx 0.74$. This is slightly lower than the value 
estimated for the chiral Ising universality class, expected to apply for the CDW transition, through various methods \cite{Janssen2014,HerbutScherer2017,Gracey:1801.01320}, but the statistical error of our result is at least on the order of $\sim 10 \%$. In any case, these results appear unphysical, since, at the very least along the $V=0$ line, the presence of CDW order is ruled out by energy balance arguments as well as by numerous other studies \cite{Assaad:1304.6340,Herbut:cond-mat/0606195,Semenoff:1204.4531}. To save space, we do not present any additional figures for these simulations. Instead, below we demonstrate that this counter-intuitive behavior is related to the violations of ergodicity in the massless HMC simulations with a single Hubbard field, i.e. at $\alpha=1$. 

 \begin{figure*}[h!tpb]
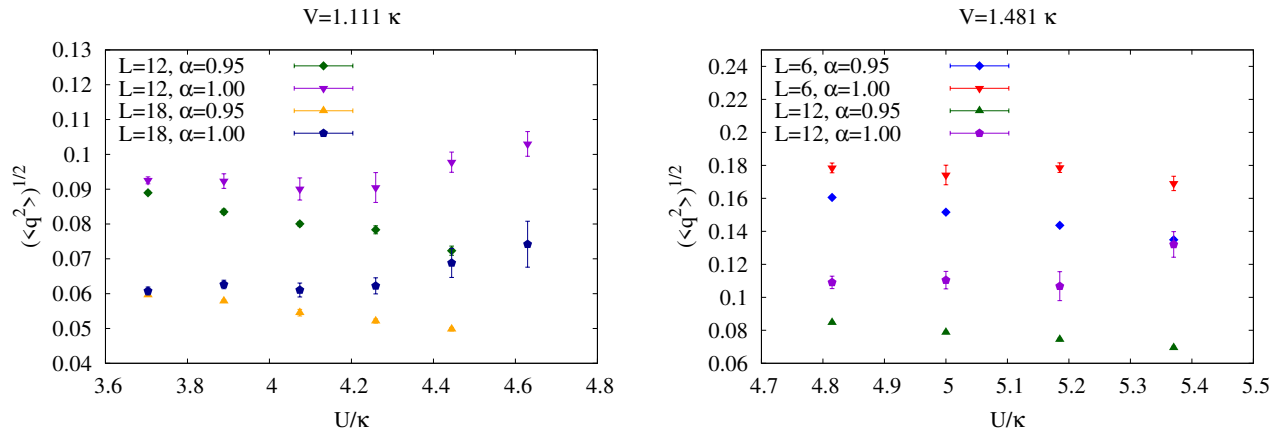

  \includegraphics[width=0.48\textwidth]{{{charge_twofield_comparison_L_12_18_V3.00}}}
  \includegraphics[width=0.48\textwidth]{{{charge_twofield_comparison_L_6_12_V4.00}}}\\
 \caption{Comparison of $\sqrt{\langle q^2 \rangle}$ as obtained from simulations with one ($\alpha=1.0$) and two ($\alpha=0.95$) Hubbard fields.}
    \label{fig:charge_twofield_comparison}
\end{figure*}

In Subsection~\ref{subsec:sdw} we 
discussed that simulations at $\alpha=0.95$ can be expected to be ergodic for every $U-V$ point considered in this work (neither $\langle q^2 \rangle$ nor $\langle S_i^2 \rangle$ depended significantly on $\alpha$ when $\alpha<0.99$ in our test cases). We now would like to further quantify the difference between ergodic and non-ergodic simulations for charge observables. Fig.~\ref{fig:charge_twofield_comparison}
shows the $U$ dependence of $\sqrt{\langle q^2 \rangle}$, obtained from simulations with $\alpha=0.95$ for the lines $V = 1.111\kappa$ and $V = 1.481\kappa$ and compares these results to the case $\alpha=1$. Unlike for $\sqrt{\langle S_i^2 \rangle}$, we observe a qualitative change: The ergodic simulations show a downward trend of $\sqrt{\langle q^2 \rangle}$ when $U$ is increased, which is lost in simulations with one Hubbard field. The ergodic and non-ergodic results drift further apart as $U$ becomes larger and in particular as we enter the SDW phase (e.g. at $U_c\approx 4.2 \kappa$ for $V= 1.111 \kappa$). Our general conclusion here is that charge is much more strongly affected than spin, by the ergodicity violations of the massless HMC simulations with a single Hubbard field. While we observed only small quantitative effects on the spin observables above, the ergodic two-field simulations here clearly allow to identify the apparent CDW order as an artifact due to these ergodicity violations.
 
\begin{figure}[h!tpb]
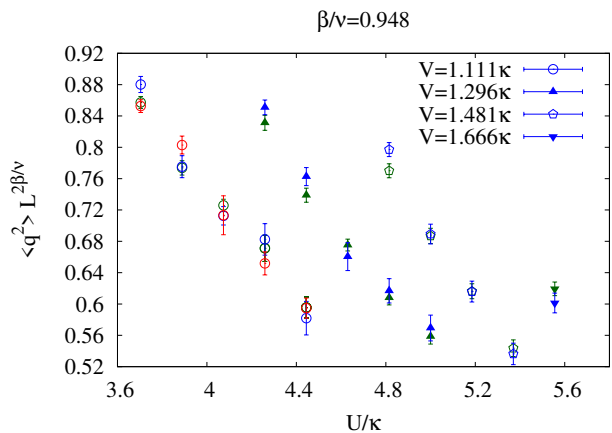

  \includegraphics[width=0.48\textwidth]{{{charge_scaling_T0.0625_twofields}}}
    \caption{Test for intersection points and finite-size scaling in $\langle q^2 \rangle$ with two Hubbard fields at $\alpha=0.95$ by comparing different lattice sizes: $L=6$ (green), $12$ (blue) and $18$ (red).}
    \label{fig:twofields_charge}
  \end{figure}

In Fig.~\ref{fig:twofields_charge} we plot $\langle q^2 \rangle L^{2\beta/\nu}$ as a function of $U$ 
for $V=1.111\kappa$, $V=1.296\kappa$, $V=1.481\kappa$ and $V=1.666\kappa$ at $\alpha=0.95$. For $V=1.111\kappa$ we show data from lattice sizes $L=6,12,18$, while for the remaining data sets results from $L=6,12$ are shown. By choosing $\beta/\nu=0.948$ we can collapse all data points of each line in the $U-V$ plane onto a single line with a very good precision. This indicates that for all our points the expectation value $\langle q^2 \rangle$ approaches zero as $\langle q^2 \rangle \sim L^{-2\beta/\nu}$ in the thermodynamic limit $L \to \infty$. Furthermore, $\langle q^2 \rangle$ decreases when $U$ is increased, in stark contrast to the non-ergodic $\alpha=1.0$ results. Thus when the complexification of the Hubbard-Stratonovich fields enables the HMC algorithm to sample the whole phase space, signatures of the CDW order appear to be just artifacts of previous non-ergodic formulation. 

\section{Conclusion and Outlook}
\label{sec:Conclusion}

We have carried out a detailed study of the SDW and CDW orders in the extended Hubbard model on the hexagonal graphene lattice with nearest-neighbour hopping and on-site and nearest-neighbour interactions $U$ and $V$. We were able to explore the region of the $U-V$ plane with $V < U/3$ and $U \lesssim 6 \kappa$. The Hybrid-Monte-Carlo algorithm which we have used becomes inapplicable for $V \geq U/3$ simulations because of a sign problem, and alternative simulation methods are required. 

We have been able to clearly identify the line of the phase transition between the semimetal phase and the gapped antiferromagnetic SDW phase, which starts at $U/\kappa = 3.9 \pm 0.04$ at $V = 0$, in agreement with the results of \cite{Assaad:1304.6340}, and bends towards larger values of $U$ as $V$ is increased. The phase transition line goes at least all the way up to the line $V = U/3$. An interesting open problem is whether it continues even to $V > U/3$. We obtained strong numerical evidence that the entire phase boundary is characterized by the same critical behavior, with a critical exponent $\beta/\nu=0.936\pm 0.022$. This is consistent within errors with the chiral Heisenberg Gross-Neveu universality class in three spacetime dimensions \cite{Gracey:1801.01320,Knorr:2018,HerbutScherer2017}. Along the $V = U/3$ line we have verified finite-size scaling with a universal scaling function for the squared spin per sublattice and estimated the correlation length exponent $\nu\approx1.162$, which further strengthens the case that this Gross-Neveu scaling persists all the way up to the $V = U/3$ line. In particular we find no evidence of multicritical or triple points in this region below $V=U/3$.  

On the other hand, our simulations suggest that charge-ordered CDW phase is absent in the region with $V < U/3$. As we have found out, the supposed signatures of the CDW phase reported in our previous work \cite{Buividovich:16:4} were the artifacts of a non-ergodic HMC algorithm which was not able to penetrate through the manifolds where the fermion determinant is zero. Similar to topology freezing in lattice QCD simulations, these manifolds are potential barriers for the molecular dynamics. The freedom of performing the Hubbard-Stratonovich transformation has allowed us to efficiently circumvent this problem. We should point out that earlier attempts to solve these issues by introducing a ``geometric mass'' (where lattice sizes are not multiples of three, so that the Dirac points do not fall on the discrete set of lattice momenta) proved to be unfruitful. 

We cannot rule out phase coexistence at exactly $V=U/3$. In this case we would expect some residual finite-size effects for points close to the line. We see no evidence for this however in Fig.~\ref{fig:twofields_charge}, where the effect should be strongest for the smallest $U$ values of each line at constant $V$. Phase coexistence at $V=U/3$ is expected in the strong coupling limit, based on energy balance arguments, so simulations at much larger values of $U$ and $V$ might be necessary to reveal a multicritical point along or close to this line. To move closer to $V=U/3$
requires simulations with values of $\alpha$ closer and closer to $\alpha = 1$ which eventually reintroduces the ergodicity problems. 

Lastly we should point out that, while simulations at $V\geq U/3$ would in principle be possible with other methods such as BSS DQMC, theses typically then suffer from a genuine fermion sign problem. At least along the $U=0$ line at finite $V$ this fermion sign problem can be avoided by exploiting a special type of time-reversal symmetry in a representation using Majorana fermions \cite{PhysRevB.91.241117,PhysRevLett.117.267002}.
This Majorana time-reversal symmetry also appears to be the reason why algorithms utilizing fermion bags \cite{PhysRevB.89.111101,PhysRevD.96.114502} or meron clusters \cite{PhysRevLett.83.3116} can be applied in such cases. 

In Ref. \cite{Ulybyshev:1712.02188} it was explicitly demonstrated that the number of relevant Lefshetz thimbles, which characterizes the severity of the sign-problem, depends on the exact form of Hubbard-Stratonovich transformation used. In particular, it was shown that switching to a non-Gaussian representation of the interaction term leads to improvements for the repulsive Hubbard model on small lattices. 

Moreover, we are currently in the process of implementing a generalized density of states method \cite{Garron2017,PhysRevD.90.094502,Langfeld2016}, which enables exponential error suppression when calculating the histogram of the phase of the fermion determinant and thus tremendously improves reweighting, for the Hubbard model at finite charge density. Extending this to $V\geq U/3$ in combination with a suitable formulation for DQMC is another possibility for future work. 

\begin{acknowledgments}
This work was supported by the Deutsche Forschungsgemeinschaft  (DFG) under grants BU 2626/2-1 and SM 70/3-1.
P.~B.~is also supported by the S.~Kowalevskaja Award from the A.~{von~Humboldt} foundation,  M.~U.~is also supported by the DFG grant AS120/14-1 and D.~S.~is also supported by the Helmholtz International Center for FAIR within the LOEWE initiative of the  State of Hesse. Calculations were carried out on GPU clusters at the Universities of Giessen and Regensburg. We thank S.~Beyl, F.~Goth and F.~Assaad for helpful discussions.
\end{acknowledgments}

\bibliographystyle{apsrev4-1}
\bibliography{Buividovich,Smith}

\appendix

\section{Schur complement solver}
 \label{apdx:Schursolver}
 In HMC simulations of fermion systems with two particle flavors (corresponding to spin-orientations in this work) it is commonplace to represent  $\det\lr{M M^{\dag}}$ as a Gaussian integral over ``pseudo-fermion'' fields $Y$
 \begin{equation}
\label{eq:stoch_det}
\det M^\dag M = \int d\bar{Y} dY \, e^{-\bar{Y} (M^\dag M)^{-1} Y}~.
\end{equation}
This representation requires repeated solutions of linear systems of the form $M X = Y$, $M^{\dag} X = Y$ or $M M^{\dag} X = Y$, which in practice is the most time consuming part of HMC simulations (up to $99 \%$ of CPU time). Typically iterative solvers, such as preconditioned Conjugate Gradient, GMRes and BiCGStab algorithms are used (in fact, the utility of a GMRes solver in simulations of the hexagonal Hubbard model was recently demonstrated \cite{Krieg:2018}), but these are efficient only for well-conditioned sparse matrices. Similar solutions are also required for the computation of Green functions, in terms of which we express physical observables. 

 In this work, we use a novel non-iterative solver based on the Schur complement, which takes the special band structure of (\ref{eq:fermionmatrix}) into account. Despite a cubic scaling with the number of lattice sites this solver outperforms iterative methods even on large lattices, as the number of operations is independent of the condition number of the matrix. Round-off errors are the only source of inaccuracy (solutions would be exact for an infinite floating point precision) and the residual is typically much smaller than for iterative solvers. To make the paper self-contained, in this Appendix we briefly describe this solver. For a much more extensive discussion, a detailed study of its efficiency in comparison with iterative methods and a pseudo-code for the algorithm, see Ref. \cite{Buividovich:18:1}.
 
 Consider that (\ref{eq:fermionmatrix}) has the structure
 \begin{equation}
\label{eq:M}
 M =
 \begin{pmatrix}
 I           & D_1 &     &        &                \\
             & I   & D_2 &        &                \\
             &     & \ddots & \ddots &                \\
             &     &        & I      &  D_{2N_\tau-1}    \\
 D_{2N_\tau}    &     &        &        &    I           \\
\end{pmatrix} ,
\end{equation}
 where the blocks $D_i$ are $N_s \times N_s$ matrices, where $N_s$ is the total number of spatial lattice sites. The Schur solver works for any matrix of this form, independent of the exact choice of $D_i$. In particular, the $D_i$ do not have to be sparse.\footnote{
 In practice, we have found that many elements of the
 non-sparse matrix (\ref{eq:fermionmatrix}) are numerically very small,
 (of order $10^{-5}$ and  smaller), and can be set
to zero without introducing any noticeable error in the results of Monte-Carlo simulations. This allows to use sparse linear algebra to further speed up the algorithm even for the exponential representation.}  In this work, all even blocks are diagonal matrices of the form
\begin{equation}
\label{eq:D2k}
 D_{2 k} = \pm
 \textrm{diag}\left( e^{i \phi^k_1}, 
 \ldots , e^{i \phi^k_{N_s}} \right)~,
 \end{equation}
where we take the plus sign for $k=N_\tau$ and the minus sign otherwise, 
while all odd blocks are non-diagonal matrices given by
 \begin{equation}
\label{eq:D2k_1_exp}
 D_{2k-1} = -e^{-{\delta_\tau} \, \, h} ,
\end{equation}
where $h$ is the single-particle hopping matrix. 

The main idea of the Schur solver is to iteratively contract the number of Euclidean time steps until the linear system $MX=Y$ can be efficiently solved using LU factorization. The contractions make use of the Schur complement (hence the name), preserve the band structure (\ref{eq:M}) of the matrix $M$ and are fully reversible, such that a solution of the original system can then be recovered. 

Consider that the vectors $X$ and $Y$ can also be rewritten in terms of blocks of size $N_s$
\begin{equation}
  \label{eq:v_blocks}
  X =
  \begin{pmatrix}
    X_1 \\ \vdots  \\  X_K
  \end{pmatrix}
  , \qquad
  Y =
  \begin{pmatrix}
    Y_1 \\ \vdots  \\  Y_K
  \end{pmatrix}
  ,
\end{equation}
where $K=2N_\tau$ for the full (uncontracted) system. At each iteration, $K$ will decrease as $K_{l+1} = \lceil K_l/2 \rceil$ where $\lceil x \rceil$ is the ceiling function.

The first step now is to apply a permutation of elements $P_K$ to the linear system:
\begin{equation}
MX=Y ~~\to ~~ (P_K M P_K^\dagger) (P_K X) = (P_K Y)~.
\end{equation}
The permutation is defined such that it mixes upper and lower halves of the vectors, i.e.
\begin{equation}
  \label{eq:perm}
  P_K X =  P_K
  \begin{pmatrix}
    X_1 \\ \vdots  \\ X_K
  \end{pmatrix} 
  =
  \begin{pmatrix}
    X_1 \\ X_{K/2+1} \\ X_2\\ X_{K/2+2} \\ \vdots  \\ X_{K/2} \\ X_K
  \end{pmatrix} 
  \equiv \overline{X}.
\end{equation}
When acting on the matrix $M$, the permutation yields
\begin{equation}
  \label{eq:perm_matrix}
 P_K M P_K^\dagger = \begin{pmatrix}
    I & R \\
    Q & J \\
  \end{pmatrix} \equiv \overline{M}~,
\end{equation}
where $I,J,R,Q$ are blocks of size $N_s K/2$. $R$ and $Q$ are given by
\begin{equation}
  \label{eq:blocks}
  R = \textrm{diag} \left(
    D_1,  D_3, \ldots,     D_{K-1} 
    \right),
\end{equation}
and
\begin{equation}
  \label{eq:Q_block}
  Q =
  \begin{pmatrix}
    0 & D_2 &        &   &   \\
      & \ddots    & \ddots &   &   \\
    &               &   \ddots      &  D_{K-2}     \\
    D_K      &       &        &  0  \\
  \end{pmatrix}~.
\end{equation}
At the first iteration (and in general for even $K$) $J\equiv I$.

To proceed, we now split the permutated vectors into upper and lower halves
\begin{equation}
  \label{eq:xy_bar}
  \overline{X} =
  \begin{pmatrix}{U_X} \\ {L_X}
  \end{pmatrix} , \quad \overline{Y}=
  \begin{pmatrix}{U_Y} \\ {L_Y}
  \end{pmatrix} ,
\end{equation}
where each half contains $K/2$ blocks of size $N_s$. The linear system $\overline{M}\,\overline{X}=\overline{Y}$
 takes the form
\begin{equation}
  \label{eq:lin_sys_new1}
  \begin{cases}
    \phantom{Q} U_X + R L_X =  U_Y , &   \\
    Q U_X + J L_X           =  L_Y .  &   \\
  \end{cases}
\end{equation}
Using the first equation we can now eliminate $U_X$ from the second equation and obtain
\begin{equation}
  \label{eq:lin_sys_new}
  \left( J - Q R \right) L_X = L_Y - Q U_Y.
\end{equation}
Once we solve this equation and find $L_X$, the upper part $U_X$ immediately follows from the first equation of (\ref{eq:lin_sys_new1}). Thus, we effectively have reduced the size of the system we must solve by a factor of two. 

A crucial point here is that the matrix $\left( J - Q R \right)$, which is the Schur complement of $\overline{M}$, has exactly the same block structure as the original matrix $M$
\begin{eqnarray}
  \label{eq:M_next_iter}
  \left( J - QR \right)
  =
  \begin{pmatrix}
    I                   & \tilde{D}_1 &             &        &                       \\
                        & I           & \tilde{D}_2 &        &                       \\
                        &             & \ddots      & \ddots &                       \\
                        &             &             & I      & \tilde{D}_{\tilde{K}-1} \\
    \tilde{D}_{\tilde{K}} &             &             &        & I                     \\
  \end{pmatrix}~,
\end{eqnarray}
with $\tilde{K}=K/2$, $\tilde{D}_k = - D_{2k} D_{2k+1}$ for $k = 1 \ldots \tilde{K}-1$
and $\tilde{D}_{\tilde{K}} = - D_K D_1$. We can thus repeat the same steps as above to iteratively shrink the system, with
the following substitution:
 \begin{eqnarray}
  \label{eq:full_iter_step}
  K & := & K/2, \nonumber \\
  M & := & J -Q R, \nonumber \\
  X & := & L_X, \nonumber \\
  Y & := & L_Y - Q U_Y .
\end{eqnarray}

In the case of odd $K$ we must artificially increase the size of the system $MX=Y$ by the block size $N_s$. By doing so, we obtain
\begin{eqnarray}M' =
  \begin{pmatrix}
    I & 0       \\
    0 & M
  \end{pmatrix},
  X' =
  \begin{pmatrix}
    0 \\ X
  \end{pmatrix},
  Y' =
  \begin{pmatrix}
    0 \\ Y
  \end{pmatrix},
  \end{eqnarray}
  and $K'=K+1$. The permutation of $M'$ now leads to
\begin{equation}
  \label{eq:matrix_bar_prime}
   \overline{M}'
  =
  \begin{pmatrix}
    I        & R' \\
    Q' & J' \\
  \end{pmatrix}
  ,
\end{equation}
 with
\begin{equation}
  \label{eq:J_and_R_block_prime}
  J'=
  \begin{pmatrix}
    I \\
    & \ddots \\
    D_{K} &   & I \\
  \end{pmatrix}
  ,\quad
  R' = \textrm{diag}\left(0,D_2,D_4, \ldots, D_{K-1}
  \right)
  ,
\end{equation}
and
\begin{equation}
  \label{eq:Q_block_prime}
  Q' =
  \begin{pmatrix}
    0 & D_1 &        &   &   \\
      & \ddots    & \ddots &   &   \\
    &        & \ddots & D_{K-2} \\
    &        &        & 0               \\
  \end{pmatrix}
  .
\end{equation}
The Schur complement $\lr{J' - Q'R'}$ again has the same structure as $M$, with $\tilde{K}=K'/2$, $\tilde{D}_k =- D_{2k-1}  D_{2k}$ for  $k = 1 \ldots \tilde{K}-1$ and  $\tilde{D}_{\tilde{K}} = D_K$.

In principle one could iterate the above procedure until Euclidean time is fully contracted and a matrix of the form $I - \prod^{2N_\tau}_{k=1} D_k$ is obtained. The final system can then be solved using $LU$ factorization. In practice, already for reasonably low temperatures, the above fully contracted matrix turns out to be extremely ill-conditioned and affected by numerical round-off errors. For this reason it is advantageous to stop after a smaller number of contractions $l_\text{max}$ (see Ref. \cite{Buividovich:18:1} for further discussions). The solution $X^{(l_\text{max})}$ is then obtained in the last iteration, after the LU decomposition of the matrix $M^{(l_\text{max})}$. Subsequently, we can revert all iterations using the relations
\begin{equation}
  \label{eq:X_iter_back}
  X^{(l-1)} = P_{l-1}
  \begin{pmatrix}{U_Y}^{(l-1)}- R^{(l-1)}  X^{(l)} \\ X^{(l)}
  \end{pmatrix}
  \end{equation}
  and reconstruct the solution of the original system. For steps in which blocks of size $N_s$ were added during the contraction we must remove them when applying (\ref{eq:X_iter_back}).

\begin{figure*}[h!tpb]
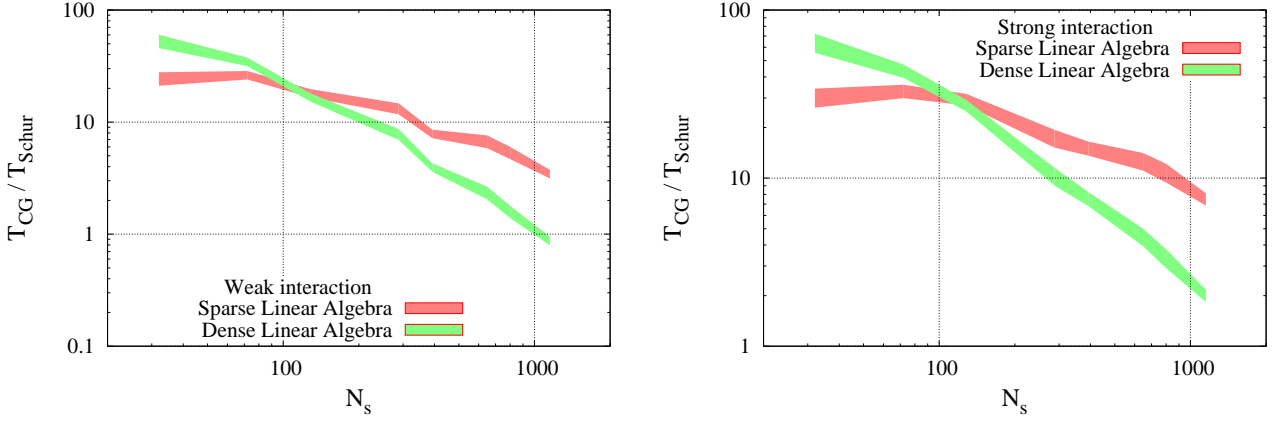

  \includegraphics[width=0.48\textwidth]{{{Ns_weak}}}
  \includegraphics[width=0.48\textwidth]{{{Ns_strong}}}\\
 \caption{Comparison of CPU runtimes of Conjugate Gradient ($T_\textrm{CG}$) and Schur solver ($T_\textrm{Schur}$) in weak-coupling (left, $U = 0.87 U_c$) and strong-coupling (right, $U = 1.07 U_c$) phases of the hexagonal Hubbard model at $N_\tau = 128$ with on-site interactions and linearized (sparse) Fermion matrix. Results are shown for linear algebra packages optimized for sparse and for dense matrices respectively. The Schur solver outperforms CG on lattice sizes up to at least $N_s=1000$ in all cases. }
\label{fig:Schur_comparison}
\end{figure*}

Finally, while the biggest strength of the Schur solver is the solution of dense systems, we would like to briefly comment on the use of this solver for matrices with initially sparse blocks $D_i$ in (\ref{eq:M}), such as the linearized Fermion operator discussed in Sec. \ref{sec:Setup}. In this case the number of floating-point operations for the solution of $MX=Y$ can be estimated as
\begin{equation}
 N_{op} 
 = 
 \sum_{l=1}^{l_\text{max}} \, N_l^2 \, N_s \, \frac{N_\tau}{2^l}
 + 
 N_{LU} ,\label{eq:scaling}
\end{equation}
where $l_{\text{max}}$ is the total number of contractions, which is limited either by $log_2\lr{N_\tau}$ or due to the accumulation of round-off errors. Here we have assumed for simplicity that $N_\tau = 2^m$ with some positive integer $m$ (the conclusions below are not changed substantially for general $N_\tau$).

$N_l$ is the number of non-zero elements in each column (row) of the blocks $D^{(l)}_i$ at the $l$-th iteration. $N_l$ grows with $l$ as
\begin{equation}
\label{eq:Nl_def}
 N_l =
 \begin{cases}
   A \, {l}^d, & A \, {l}^d < N_s , \\
   N_s,        & A \, {l}^d > N_s , \\
 \end{cases} 
\end{equation}
where $d$ is the number of spatial lattice dimensions and $A$ is a numerical pre-factor which depends on the details of the theory, such as the number of Fermion components and the number of nearest neighbors on a lattice of a given type. $N_{LU}$ is the number of floating-point operations required for the LU decomposition, which scales with $N_s$ and $N_\tau$ as 
\begin{equation}
\label{eq:NLU_scaling}
 N_{LU} \sim \lr{N_s \, \frac{N_\tau}{2^{l_\text{max}}}}^3 .
\end{equation}

 Fig. \ref{fig:Schur_comparison} shows a comparison of the CPU runtimes of the Schur solver and a standard CG solver for the Hubbard model with on-site interactions only in the strong-coupling ($U = 1.07 U_c$) and weak-coupling ($U = 0.87 U_c$) phases at $N_\tau = 128$. As the initially sparse blocks $D^{l}_k$ become denser after each contraction, it can be advantageous to use linear algebra packages optimized for dense matrices for the matrix operations. The figure displays the comparison for both dense and sparse linear algebra. 
 As expected, the largest speedup is achieved for smaller lattices. In this case the use of dense linear algebra is also extremely beneficial.

  The overall conclusion is that in the strong-coupling phase the Schur solver is faster than CG even for lattices with $N_s = 1000$.  When sparse linear algebra routines are used, the speed-up is at least a factor of ten and depends rather weakly on the lattice size. A rough extrapolation suggests that in the strong-coupling phase the Schur complement solver outperforms CG for lattice sizes up to at least $N_s \sim 10^4$. In the weak-coupling phase the speed-up is smaller but also significant. Again, a rough extrapolation suggests that in this regime the Schur solver outperforms CG up to about $N_s \sim 10^3 \ldots 10^4$.

\section{Expressing observables in terms of Green functions}
 \label{apdx:Observables}
 
We express each observable in terms of the full fermion equal-time Green function $g(x,y)=\langle\hat{a}_x \hat{a}^{\dagger}_y\rangle=M^{-1}_{x,t,y,t}$, which is computed for every lattice configuration. For the $S_1$, $S_2$ components of the squared spin per sublattice we obtain
\begin{align}
\langle S_{1,2}^2 \rangle  = &\frac{1}{4L^4} \Big\{ \sum_{x\in A} (1-2 \mbox{Re} \, g(x,x)) \notag\\ &+\sum_{ {x,y \in A}} ( |g(x,y)|^2 + |g(y,x)|^2 ) \Big\} ,
\end{align}
and
\begin{align}
&\langle S_3^2 \rangle 
= \frac{1}{4L^4} \Big\{ \sum_{x \in A}  (1-2 \mbox{Re} \, g(x,x)+2 |g(x,x)|^2 ) \notag\\
&\quad\quad+\sum_{x,y\in A; x\neq y}  \big\{ 1 + 2 \mbox{Re} \, \big[ g (x,x) g(y,y)- g(x,y) g (y,x)  \notag\\
&\quad\quad+ g(x,x)^{*} g(y,y) -2g(x,x) \big] \big\} \Big\}.
\end{align}
Similarly, for the squared charge per sublattice we obtain:
\begin{align}
\langle q^2 \rangle
 =&  \frac{2}{L^4}\Big\{\sum_{x \in A} ( \mbox{Re} \, g(x,x)- |g(x,x)|^2 )\notag\\
 &\quad\quad+ \sum_{x,y\in A; x\neq y} 
\mbox{Re}\big[ g(x,x)g(y,y)
\notag\\&\quad\quad- g(y,x)g(x,y) - g(x,x)g(y,y)^* \big]\Big\}.
\end{align}
Note that the sums here run over sublattice ``A'' only. To recover 
eqs. (\ref{eq:squarespin}) and 
(\ref{eq:squarecharge}) one should sum also over sublattice ``B'' and then add both results together.

For the components $m_1$, $m_2$ of the mean magnetization we obtain
\begin{align}
\langle m_{1,2}^2 \rangle  = &\frac{1}{4L^4} \Big\{ \sum_{x} (1-2 \mbox{Re} \, g(x,x)) \notag\\ &+\sum_{ {x,y}} ( |g(x,y)|^2 + |g(y,x)|^2 )P(x,y) \Big\} ,
\end{align}
where $P(x,y)=1$ if $x$ and $y$ belong to the same sublattice and $P(x,y)=-1$ otherwise. The expression for $m_3$ is
\begin{align}
&\langle m_3^2 \rangle 
= \frac{1}{4L^4} \Big\{ \sum_{x}  (1-2 \mbox{Re} \, g(x,x)+2 |g(x,x)|^2 ) \notag\\
&\quad\quad+\sum_{x,y; x\neq y}  \big\{ 1 + 2 \mbox{Re} \, \big[ g (x,x) g(y,y)- g(x,y) g (y,x)  \notag\\
&\quad\quad+ g(x,x)^{*} g(y,y) -2g(x,x) \big] \big\} \Big\},
\end{align}
which differs from $\langle S_3^2 \rangle$ only by the range of the sums.

\end{document}